\documentclass[twocolumn]{openjournal}
\usepackage{graphicx} 
\usepackage{multirow}

\usepackage[colorlinks,linkcolor=blue,citecolor=blue,urlcolor=blue ]{hyperref}
\usepackage[utf8]{inputenc}
\usepackage{float}
\usepackage{xcolor}
\usepackage{ulem}
\usepackage[T1]{fontenc}

\def\msun{M$_{\odot}$}
\def\logm{log (M$_{\star}$/M$_{\odot}$)}

\def\sigh2{$\Sigma_{\rm H_2}$}

\def\kms{km~s$^{-1}$}

\def\mhi{M(HI)}
\def\fgas{$f_{\rm HI}$}

\begin{document}
\title{Low redshift post-starburst galaxies host abundant HI reservoirs\vspace{-1.5cm}}

\author{Sara L. Ellison$^1$}
\author{Qifeng Huang$^{2,3}$}
\author{Dong Yang$^{2,3}$}
\author{Jing Wang$^2$}
\author{Vivienne Wild$^4$}
\author{Ben F. Rasmussen$^1$}
\author{~~~~~~~~~~~~~~~~~Maria Jesus Jimenez-Donaire$^{5,6}$}
\author{Kate Rowlands$^{7,8}$}
\author{Scott Wilkinson$^1$}
\author{Toby Brown$^9$}
\author{Ho-Hin Leung$^{4,10}$}
\thanks{$^*$E-mail: sarae@uvic.ca}
\affiliation{$^1$ Department of Physics \& Astronomy, University of Victoria, Finnerty Road, Victoria, BC V8P 1A1, Canada\\
  $^2$ Kavli Institute for Astronomy and Astrophysics, Peking University, Beijing 100871, People’s Republic of China\\
  $^3$ Department of Astronomy, School of Physics, Peking University, Beijing 100871, People’s Republic of China\\
  $^4$ School of Physics and Astronomy, University of St Andrews, North Haugh, St Andrews, KY16 9SS, U.K.\\
  $^5$ Observatorio Astronómico Nacional (IGN), C/Alfonso XII 3, 28014, Madrid, Spain\\
  $^6$ Centro de Desarrollos Tecnológicos, Observatorio de Yebes (IGN), 19141 Yebes, Guadalajara, Spain\\
  $^7$ AURA for ESA, Space Telescope Science Institute, 3700 San Martin Drive, Baltimore, MD 21218, USA\\
  $^8$ William H. Miller III Department of Physics and Astronomy, Johns Hopkins University, Baltimore, MD 21218, USA\\
  $^9$ National Research Council of Canada, Herzberg Astronomy and Astrophysics Research Centre, 5071 W. Saanich Rd. Victoria, BC, V9E
  2E7, Canada\\
  $^{10}$ SUPA, Institute for Astronomy, University of Edinburgh, Royal Observatory, Edinburgh EH9 3HJ, UK
}

\begin{abstract}

Studying the gas content of post-starburst (PSB) galaxies can provide valuable clues regarding the process of fast quenching. Although previous works have studied the molecular gas content of PSBs, only a handful of HI measurements exist.  Here, we present new Five hundred metre Aperture Spherical Telescope (FAST) 21cm observations of 44 PSBs, leading to 43 detections or sensitive upper limits of HI, which we combine with 25 archival measurements, for a total sample of 68 PSB \mhi\ measurements.     HI is detected in 57/68 galaxies, with HI masses ranging from \mhi\ $\sim10^{8.5}$ up to $10^{10}$ \msun\ and gas fractions (\fgas\ = \mhi/M$_{\star}$) from a few percent up to almost 30 percent.  Post-starbursts therefore retain ample atomic gas reservoirs, despite no longer forming stars.  By comparing with a stellar mass-matched sample of star-forming galaxies in xGASS, we find that PSBs have, on average, gas fractions lower by $\sim$ 0.2-0.4 dex, consistent with a mild reduction compared with their progenitor population.  However, PSBs show a diversity of HI properties; about half have HI gas masses within the expected scatter of the star-forming population with the remaining 50 per cent up to a factor of 10 more gas-poor.  Compared with galaxies in the green valley, about two thirds of PSBs have gas fractions within the expected range, with the remaining third up to a factor of 10 more gas-rich.  Our results demonstrate that quenching in PSBs is not the result of wholesale removal of the atomic gas reservoir and that the population has atomic gas fractions that span the range from star-forming to green valley galaxies.  We find no correlation between HI gas mass and time since burst; even galaxies a Gyr past their burst can remain HI-normal. The significant gas reservoirs remaining in many PSBs leaves open the possibility for future rekindling of star formation.

\end{abstract}
\maketitle

\section{Introduction} \label{intro_sec}

The bimodality of galaxy properties can be traced back to the early observations by Hubble, who showed that galaxies can be broadly classified into two types: spirals and ellipticals (Hubble 1926).  After a century of observations, we now know that many galaxy properties, not just morphologies, show bimodal distributions; galaxies tend to be either blue, disk-dominated and actively forming stars, or red ellipticals with very low rates of star formation (e.g. Strateva et al. 2001; Baldry et al. 2006; Wuyts et al. 2011).  Large spectroscopic galaxy surveys have also demdonstrated this bimodal behavior in the distribution of star formation rates (SFRs) with stellar mass.  Whereas star-forming galaxies form a tight `main sequence', the `quenched' population has SFRs that are lower by at least an order of magnitude at fixed stellar mass (e.g. Brinchmann et al. 2004; Elbaz et al. 2007; Salim et al. 2007; Wyder et al. 2007; Schawinski et al. 2014).

\smallskip

Understanding the mechanisms that trigger the migration of a galaxy from the star forming main sequence to the quenched population has been a major focus of extra-galactic research in the last 15 years (e.g. Peng et al. 2010; Bluck et al. 2014, 2016; Omand et al. 2014; Woo \& Ellison 2019; Brownson et al. 2022 to name but a few).  There are numerous possible causes for such a transition.  For example, star formation will necessarily cease if all/most of the interstellar gas reservoir has been consumed through past star formation.  Alternatively, the gas may be forcibly removed through outflows that result from energetic events such as active galactic nuclei (AGN) or starbursts.   More gentle feedback can also play a role, by preventing halo gas from reaching the disk and forming new stars. Gas may also be removed by external forces, such as ram pressure stripping in clusters.  Finally, the gas reservoir may remain largely intact, but be unfavorable to the formation of stars, e.g. due to the stabilizing effect of a bulge, a highly turbulent medium, or surface densities below a critical threshold (e.g. Lemonias et al. 2014; Gensior et al. 2020; Smercina et al. 2022).  The key to making observational progress in distinguishing which of these quenching pathways actually operates in the real Universe is therefore via a detailed assessment of gas, not only its presence, but also its spatial distribution, thermal structure and dynamics.  Ideally, such an assessment would be done for a population of galaxies caught in the act of quenching, allowing us to characterize the properties of the interstellar medium that accompany the shut-down of star formation.  Post-starburst (PSB) galaxies, characterized by spectral features that identify them as having had a recent history of active star formation that has subsequently been rapidly and recently quenched (see French 2021 for a review), offer such an opportunity.

\smallskip

The global molecular gas content of low redshift PSBs has been studied in numerous previous works (e.g. French et al. 2015; Rowlands et al. 2015; Alatalo et al. 2016; Yesuf et al. 2017; French et al. 2018a; Smercina et al. 2018; Yesuf \& Ho 2020) in samples whose sizes number from a few tens to over 100.  Despite some differences linked to details of the sample selection, the broad consensus from these previous works is that PSBs frequently host large reservoirs of cold gas, with molecular gas fractions ranging from that of actively star-forming galaxies to the gas-rich end of the quiescent population.  The shut-down of star formation in (low redshift) PSBs therefore seems unrelated to the wholesale removal/depletion of the molecular interstellar medium.  Instead, it has been proposed that increased turbulence, possibly related to the high incidence of mergers amongst PSBs (e.g.  Pawlik et al. 2018; Sazonova et al. 2021; Wilkinson et al. 2022; Ellison et al. 2024), could be responsible for the reduction in star formation (Otter et al. 2022; Smercina et al. 2022).  As a result of this turbulence, the remaining diffuse molecular gas is unable to attain sufficient densities to continue forming stars, as supported by low fractions of dense gas e.g. as traced by HCN (French et al. 2018b; French et al. 2023).  Therefore, in addition to demonstrating that quenching can occur even in the presence of a substantial gas supply, these previous works also raise the possibility that star formation in these galaxies could re-ignite again in the future.

\smallskip

It is clear from these previous observations that understanding PSB evolution requires studying the many phases of gas in the interstellar medium.  Atomic gas is a vital component of this assessment.  Although not directly involved in the formation of stars (which are born from the dense molecular phase), from a cosmic perspective HI dominates over H$_2$ out to at least $z\sim6$ (and likely beyond; see Peroux \& Howk 2020 for a review).  Moreover, the largely unevolving (statistical) HI content of galaxies over time (e.g. Crighton et al. 2015; Sanchez-Ramirez et al. 2016; Berg et al. 2019) implies that the atomic reservoir experiences steady replenishment, which in turn has the potential to fuel future star formation.  The presence and physical state of atomic gas in PSBs is therefore an important factor both in understanding the reason for quenching, as well as the potential for rejuvenation.

\smallskip

In contrast to the fairly extensive investigations of molecular gas in PSBs described above, it is somewhat surprising (given the relative ease of measuring HI in low redshift galaxies) that their atomic gas properties remain poorly explored.  To our knowledge, targeted HI observations of PSBs have been presented in only four previous papers: 11 PSBs in the Coma cluster by Bravo-Alfaro et al. (2001), 5 galaxies by Chang et al. (2001), 6 galaxies by Buyle et al. (2006), 11 galaxies by Zwaan et al. (2013).  One further paper, by Li et al. (2023), cross-matched their PSB sample with the HI-MaNGA survey  (Stark et al. 2021) and found 15 detections.  With the exception of the PSBs in Coma (which seem to be uniformly gas poor) these studies detect HI in approximately 50 percent of cases.  It has therefore been suggested that PSBs frequently conserve significant atomic gas reservoirs, although there is substantial diversity in their gas fractions.

\smallskip

Despite the important finding that PSBs can harbour significant HI reservoirs, these previous studies suffer from a number of short-comings.  First, the sample sizes are obviously small.  Second, the detection limits are relatively shallow (e.g. HI mass limits of $\sim 3 \times 10^9$ \msun\ in Chang et al. 2001 and Buyle et al. 2006) and often require stacking to make comparisons (Li et al. 2023). Third, previous studies are performed to heterogenous depths, making it challenging to draw statistical statements from their compilation.  Fourth, these previous studies largely lack a rigourous comparison to a control sample observed to the same depths and with the same methods.  Therefore, whilst a qualitative statement about the presence of HI can be made, a quantitative assessment remains elusive.  Finally, a significant fraction of the galaxies in these previous works are in clusters, which is an atypical environment for low redshift PSBs (Yesuf 2022).  For example, Bravo-Alfaro et al. (2001) discuss how their results in the Coma cluster indicate a truly different evolutionary path compared with the more common field PSBs.

\smallskip

In the work presented here we aim to conduct the first statistical assessment of HI in a representative sample of low redshift PSBs.  By adopting a consistent PSB selection scheme, for a large sample, with a deep and uniform detection threshold and with a carefully controlled comparison with HI in `normal' galaxies, we can finally assess the atomic gas reservoirs of recently quenched galaxies in a robust and quantitative way.  Moreover, in a follow-up paper (Rasmussen et al., in prep) we will present molecular gas observations for the same sample.  As a result, we will ultimately be able to assess the interplay between the atomic and molecular gas phases during the epoch of rapid quenching.

\section{Data and methods}

\begin{deluxetable*}{cccccccccc}
\tablecaption{FAST observations of PSBs. \label{tab:obs}}
\tablewidth{\textwidth}
\tablehead{SDSS ObjID & R.A. & Decl. & $z_{\rm SDSS}$ & $\log M_\ast$ & $T_{\rm on}$ & $\int F_v{\rm d}v$ & rms & S/N & W85 \\
 & (deg) & (deg) & & ($M_\odot$) & (min) & ($\rm Jy~km~s^{-1}$) & (mJy) & & \kms }
\startdata
587722984440463382 & 216.55409 & 0.86058 & 0.031859 & 10.18 & 13.0 & $0.26 \pm 0.02$ & 0.25 & 19.8 & 223 \\
587726032256630848 & 198.46830 & 2.13257 & 0.030259 & 10.53 & 9.0 & $0.63 \pm 0.01$ & 0.34 & 30.2 & 309 \\
587729408621609096 & 259.53273 & 30.12902 & 0.029660 & 9.87 & 9.7 & $0.12 \pm 0.01$ & 0.20 & 9.6 & 141\\
587729652890206755 & 256.43331 & 31.41377 & 0.034796 & 9.81 & 18.7 & $0.19 \pm 0.01$ & 0.11 & 36.0 & 274 \\
587730847963545655 & 318.50226 & 0.53510 & 0.026921 & 10.18 & 6.5 & $0.79 \pm 0.02$ & 0.39 & 45.9 & 326\\
587730846889869791 & 318.67133 & -0.41098 & 0.032113 & 10.27 & 13.5 & $0.09 \pm 0.01$ & 0.22 & 10.2 & 117 \\
587731174382502294 & 319.95072 & 0.67272 & 0.034424 & 9.61 & 17.9 & $0.23 \pm 0.01$ & 0.20 & 19.0 & 525 \\
587731186735186207 & 347.26187 & 0.26689 & 0.032520 & 10.19 & 14.2 & $0.34 \pm 0.01$ & 0.16 & 48.7 & 292\\
587731512082956346 & 50.88860 & -0.43856 & 0.023876 & 9.88 & 4.0 & $<0.21$ & 0.47 & -- & -- \\
587731681190740132 & 129.18887 & 39.83509 & 0.024715 & 9.96 & 4.6 & $0.30 \pm 0.02$ & 0.33 & 46.9 & 319\\
587732482747400295 & 176.81117 & 49.18526 & 0.025601 & 9.65 & 5.3 & $0.11 \pm 0.02$ & 0.25 & 7.4 & 131\\
587732580982521898 & 169.78179 & 58.05398 & 0.032604 & 10.54 & 11.9 & $0.29 \pm 0.02$ & 0.22 & 15.3 & 214\\
587732701250519063 & 149.29786 & 5.20132 & 0.021686 & 9.94 & 2.7 & $<0.18$ & 0.39 & -- & --\\
587734622705811566 & 131.15756 & 32.90645 & 0.031543 & 10.31 & 12.5 & $0.23 \pm 0.01$ & 0.17 & 38.7 & 248 \\
587732771584671859 & 164.59071 & 9.45391 & 0.033594 & 9.77 & 16.2 & $0.19 \pm 0.01$ & 0.20 & 21.7 & 309 \\
587735349098971209 & 154.24553 & 13.39929 & 0.032541 & 9.77 & 14.2 & $0.05 \pm 0.01$ & 0.20 & 5.3 & 130 \\
587735431226130460 & 191.61184 & 50.79205 & 0.027056 & 10.56 & 5.1 & $0.22 \pm 0.02$ & 0.35 & 15.2 & 267\\
587735664773431424 & 226.15445 & 48.73879 & 0.036099 & 10.53 & 18.5 & $0.50 \pm 0.01$ & 0.17 & 71.3 & 303\\
587736541491233170 & 237.71840 & 5.32690 & 0.026031 & 9.66 & 5.7 & $1.25 \pm 0.03$ & 0.52 & 56.2 & 241\\
587738410325573774 & 150.69248 & 12.27086 & 0.021737 & 9.72 & 2.7 & $<0.19$ & 0.42 & -- & -- \\
587738615415373880 & 170.94590 & 35.44231 & 0.034067 & 10.29 & 17.1 & $0.18 \pm 0.01$ & 0.14 & 16.7 & 250 \\
587738067810189520 & 133.03863 & 64.07869 & 0.036392 & 9.55 & 22.5 & $<0.08$ & 0.18 & -- & -- \\
587739130805420150 & 206.44644 & 34.49324 & 0.034790 & 9.54 & 18.7 & $0.40 \pm 0.01$ & 0.15 & 67.2 & 216\\
587739303684931731 & 198.10781 & 34.12134 & 0.033809 & 9.89 & 16.6 & $0.14 \pm 0.01$ & 0.14 & 23.3 &119 \\
587739504477077652 & 204.01718 & 30.14107 & 0.025928 & 9.64 & 5.6 & $0.40 \pm 0.00$ & 0.24 & 16.3 & 127\\
587739166775574594 & 245.25338 & 21.16835 & 0.031025 & 10.38 & 11.7 & $0.10 \pm 0.01$ & 0.17 & 6.1 & 137 \\
587739609698140217 & 176.73898 & 32.65397 & 0.032757 & 9.61 & 14.6 & $0.04 \pm 0.01$ & 0.17 & 6.0 & 78\\
587739844856971464 & 241.64677 & 14.36721 & 0.032489 & 9.68 & 14.1 & $0.14 \pm 0.01$ & 0.23 & 18.9 & 104\\
587741722823950491 & 195.50058 & 27.78273 & 0.023876 & 9.82 & 4.0 & $<0.13$ & 0.30 & -- & --\\
587741815710744668 & 155.10063 & 21.35628 & 0.019190 & 10.02 & 1.7 & $0.36 \pm 0.02$ & 0.51 & 17.5 & 266 \\
587742576459251895 & 225.34016 & 15.24990 & 0.035373 & 9.91 & 20.0 & $0.27 \pm 0.01$ & 0.15 & 30.5 & 252\\
587742589333340203 & 245.50907 & 9.88863 & 0.032715 & 10.48 & 14.5 & $0.09 \pm 0.00$ & 0.14 & 13.4 & 280 \\
587742188827443225 & 174.95152 & 23.53247 & 0.030655 & 10.49 & 11.1 & $<0.11$ & 0.20 & -- & --\\
587742189363527850 & 173.01533 & 23.70381 & 0.032286 & 10.19 & 13.8 & $<0.10$ & 0.22 & -- & -- \\
587742616170266870 & 235.75415 & 16.98744 & 0.031407 & 9.87 & 12.3 & $0.20 \pm 0.01$ & 0.18 & 23.7 & 141\\
587742644626260055 & 240.09445 & 12.75777 & 0.034376 & 10.12 & 17.8 & $<0.07$ & 0.16 & -- & --\\
587745244159082657 & 136.46640 & 13.71746 & 0.027348 & 9.68 & 7.0 & $0.24 \pm 0.01$ & 0.23 & 11.0 & 162\\
587745403073855572 & 137.87957 & 12.14780 & 0.029704 & 10.14 & 9.8 & $<0.09$ & 0.19 & -- & --\\
588013382740279420 & 170.72341 & 51.34171 & 0.034165 & 9.85 & 17.4 & $<0.10$ & 0.21 & -- & --\\
588015508189872263 & 346.93215 & -0.83848 & 0.032532 & 9.52 & 14.2 & $1.63 \pm 0.05$ & 0.21 & 26.3 & 218 \\
588015508208156794 & 28.63643 & -0.77010 & 0.016178 & 9.66 & 0.8 & $<0.30$ & 0.67 & -- & -- \\
588016839633666164 & 123.32063 & 22.64830 & 0.022299 & 9.90 & 3.0 & $0.15 \pm 0.02$ & 0.45 & 9.2 & 106 \\
588017627780087861 & 227.22954 & 37.55827 & 0.029058 & 10.21 & 8.9 & $0.71 \pm 0.03$ & 0.30 & 29.9 & 134\\
588298664112881826 & 192.66358 & 47.93427 & 0.029111 & 9.63 & 9.0 & $0.59 \pm 0.01$ & 0.30 & 51.4 & 301
\enddata
\tablecomments{Columns are: 1 -- SDSS DR7 ObjID; 2 -- Right ascension; 3 -- Declination; 4 -- Redshift from the SDSS DR7 optical spectrum; 5 -- Log of the stellar mass from the MPA/JHU catalog; 6 -- FAST on-source integration time; 7 -- Integrated 21cm line flux or 5$\sigma$ upper limit; 8 -- RMS of the final science spectrum; 9 -- S/N of the detected 21cm line; 10 -- line width that contains 85 \% of the flux.}
\end{deluxetable*}

\subsection{Sample selection}

Two complementary PSB catalogs are used to select our parent sample.  First, in order to select traditional E+A PSBs, we used the classifications\footnote{Available in the catalog presented at http://www.phys.nthu.edu.tw/$\sim$tomo/cv/index.html} for the SDSS DR7 that followed methods developed for earlier data releases (e.g. Goto 2005, 2007).   These E+A PSBs were identified according to the following equivalent width (EW) thresholds (where positive values indicate absorption and negative values indicate emission): EW(H$\delta$)$>$ 5.0 \AA, EW([OII])$>-$2.5 \AA\ and EW(H$\alpha$)$>-$3.0 \AA.  Second, for a more inclusive sample of PSBs, we used the principal component analysis (PCA) of Wild et al. (2007), again applied to SDSS DR7, which selects galaxies with a relative excess of Balmer absorption given the age of their stellar population and ignores emission line strengths.  In addition to cuts in PCA space, we additionally apply mass dependent thresholds in Balmer decrement designed to remove dusty galaxies that may not be true post-starbursts.  Further details of the PCA selection are given in full in Wilkinson et al. (2022).

\smallskip

We began by selecting all PSBs from the above two catalogs whose redshift and stellar mass lie in the ranges $0.01 < z< 0.04$ and \logm\ $ > 9.5$.  These cuts provided a complete and representative sample of PSB galaxies whose 21cm emission lines could be detected in reasonable amounts of observing time (more details on detection thresholds are given below). A sample of 114 galaxies was thus selected.  The SDSS spectra for the 114 targets were visually inspected to ensure that they did indeed manifest strong Balmer absorption features, as expected of PSBs.  Three galaxies were rejected as a result of the visual inspection.

\smallskip

Due to the large beam size of FAST ($\sim$ 3 arcmin), it was important to reject galaxies from the parent sample if they had close companions that might contribute additional 21cm flux and hence contaminate the HI mass measurement of the PSB.  In order to discard potentially contaminating sources we select galaxies in the Dark Energy Spectroscopic Instrument (DESI) Legacy Imaging Survey that have consistent photometric ($\Delta z < 3\sigma_{z_{phot}}$, Zhou et al. 2023) or spectroscopic redshifts ($\Delta cz_{spec}<600~{\rm km~s^{-1}}$) within $5'$. Since spectroscopic masses (that are used for our final science analysis) are not available for every galaxy, for the beam contamination check we use photometrically deduced stellar masses.  Specifically, for each galaxy we estimate a stellar mass and HI gas fraction from GALEX NUV and SDSS optical photometry (Bianchi et al. 2017; Ahumada et al. 2020) based on the procedures described by Zibetti et al. (2009) and Zhang et al. (2021). From the stellar masses and gas fractions we then calculate HI flux estimates, accounting for the radially diminishing FAST beam response. Contamination is considered significant when the predicted flux ratio between the neighbours and the target exceeds 1:3. 25 galaxies were rejected as a result of the contamination checks, leaving a final sample of 86 PSBs.  Although this contamination cut removes PSBs from our sample if they have a companion within $\sim$ 100 kpc, PSBs do not seem to be prefentially associated with galaxies in this regime (Ellison et al. 2022).  Our contamination cuts also potentially bias us against dense cluster environments.  However, low $z$ PSBs are generally not in over-dense environments (Yesuf 2022).  Indeed, we checked the larger scale environments of the rejected PSBs, as measured by distances to the 5th nearest neighbour, and find that their values are consistent with the remaining PSBs in our sample.  We therefore conclude that although our final PSB sample is devoid of galaxies with very near neighbours, their gas fractions should still be representative of the general (low redshift) PSB population.

\smallskip

In order to assess whether there were existing HI mass measurements for any of the 86 galaxies in our PSB sample, we checked the ALFALFA (Haynes et al. 2018), xGASS (Catinella et al. 2018), HI-MaNGA (Stark et al. 2021) and FASHI (Zhang et al. 2024) surveys for archival observations.  Since ALFALFA and FASHI are blind surveys, no upper limits are provided in the public catalogs, so only detections are taken for targets matched to these surveys (i.e. non-detections are considered as non-observations).  Conversely, HI-MaNGA and xGASS are targeted surveys and HI mass upper limits are available in the case of non-detections.  However, since the HI-MaNGA upper limits are shallower than our target detection theshold (see below), non-detections in HI-MaNGA were considered as non-observations.  We therefore potentially have upper limits available only for xGASS archival matches.  As a result of the survey cross-matching we obtained archival HI masses for 26 out of 86 galaxies, leaving 60 PSBs requiring new observations.

\begin{figure*}
    \centering
    \includegraphics[width=\linewidth]{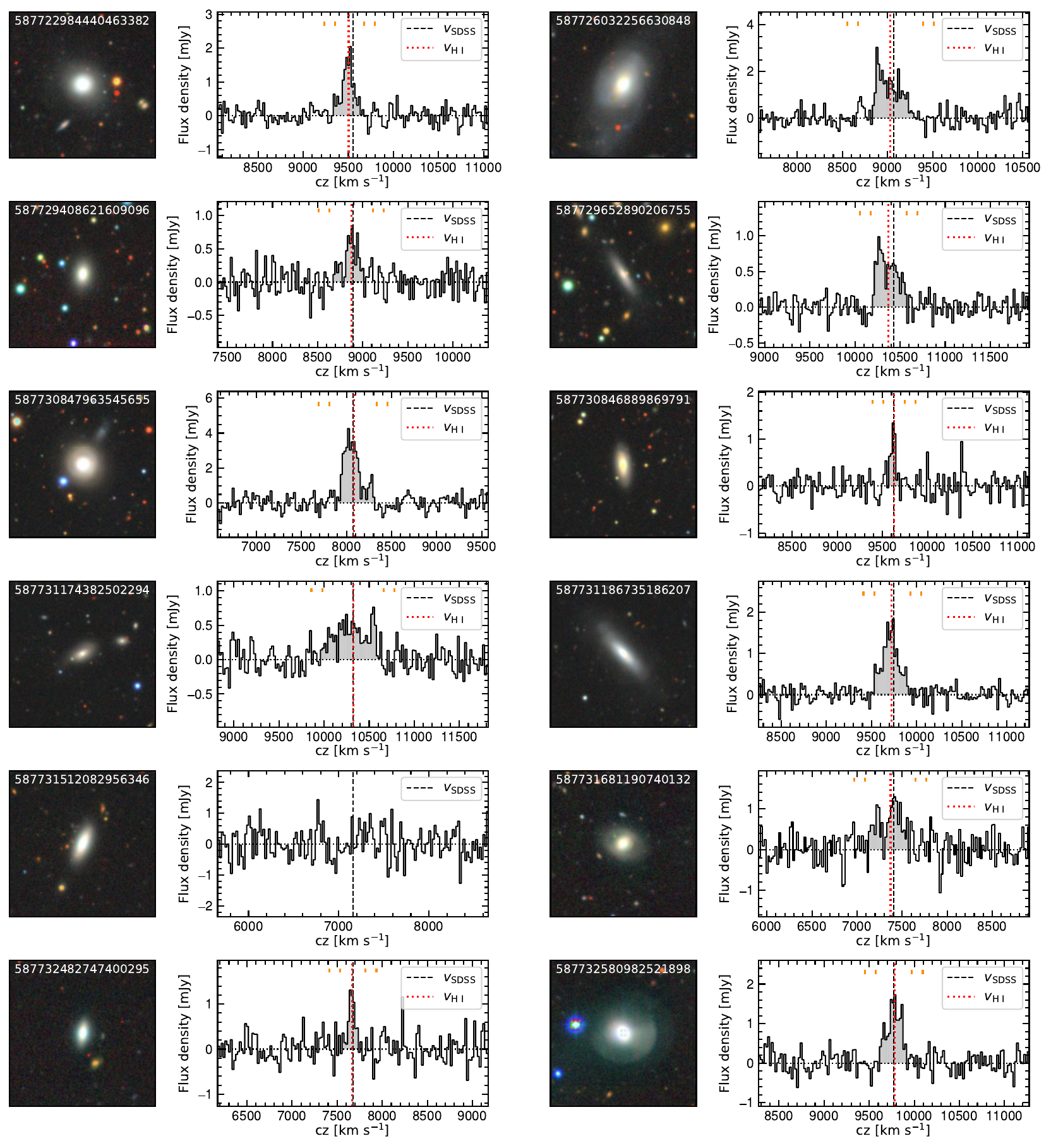}
    \caption{Legacy survey images ($1'.45\times1'.45$, half of the FAST beam size; Dey et al. 2019) and HI spectra for the first 12 galaxies in Table \ref{tab:obs}, ordered by the SDSS objID (shown in the images). The HI spectra are baseline-subtracted and binned to a channel width of $20~\rm km~s^{-1}$. The red dotted line and the black dashed line indicate the heliocentric velocity from HI spectra and SDSS, respectively. The orange dashes show the velocity range where the HI growth curve flattens. Images and spectra for the remaining galaxies in our sample are presented in the Appendix (Figure \ref{atlas_all}).  }
    \label{atlas_fig}
\end{figure*}

\subsection{Observations and data reduction}

We were granted 25 hours on the FAST telescope in its fourth observing cycle, from August 2023 until July 2024 (Proposal ID: SQB-2023-0065, PI: Ellison).  Unfortunately, 25 hours (which was less than the original request) were insufficient to observe all of the 60 targets requiring new observations down to the desired detection threshold (see below).  We therefore prioritized the lower redshift targets, imposing an effective redshift cut of $z<0.0364$ (although some of the archival observations are at $0.0364 < z< 0.04$).  Thus, of the 60 PSBs requiring new observations, only 44 were observed.  Table \ref{tab:obs} summarizes the targets observed in our FAST campaign, along with relevant observational details. 

\begin{table}
  \begin{center}
  \caption{HI gas masses for the PSB sample.  }
\begin{tabular}{ccccc}
\hline

SDSS ObjID & $z_{SDSS}$ & log M$_{\star}$ & log \mhi & HI source \\ 
           &   & (M$_{\odot}$) & (M$_{\odot}$) & \\
\hline                                                                                                      
 & & & & \\
587722984440463382   &    0.03186 &     10.18 &     9.06$\pm$0.05 &    5 \\ 
587725489988960505   &    0.02961 &      9.61 &      8.86 &    3 \\ 
587726032256630848   &    0.03026 &     10.53 &     9.4$\pm$0.04 &    5 \\ 
587728931343630410   &    0.03934 &     10.37 &      9.69 &    4 \\ 
587729408621609096   &    0.02966 &      9.87 &     8.67$\pm$0.06 &    5 \\ 
587729652890206755   &    0.03480 &      9.81 &     9.0$\pm$0.05 &    5 \\ 
587730023333232703   &    0.03564 &     10.50 &     9.89$\pm$0.05 &    1 \\ 
587730773885059094   &    0.02563 &      9.99 &     9.45$\pm$0.06 &    1 \\ 
587730847963545655   &    0.02692 &     10.18 &     9.4$\pm$0.04 &    5 \\ 
587730846889869791   &    0.03211 &     10.27 &     8.62$\pm$0.06 &    5 \\ 
587731174382502294   &    0.03442 &      9.61 &     9.08$\pm$0.05 &    5 \\ 
587731186735186207   &    0.03252 &     10.19 &     9.19$\pm$0.04 &    5 \\ 
587731886809808959   &    0.03975 &      9.89 &      9.67 &    3 \\ 
587731512082956346   &    0.02388 &      9.88 &     <8.72 &    5 \\ 
587731512619106324   &    0.02318 &     10.08 &       8.6 &    3 \\ 
587732482747400295   &    0.02560 &      9.65 &     8.51$\pm$0.09 &    5 \\ 
587732580982521898   &    0.03260 &     10.54 &     9.13$\pm$0.05 &    5 \\ 
587732701250519063   &    0.02169 &      9.94 &     <8.56 &    5 \\ 
587734622705811566   &    0.03154 &     10.31 &     9.0$\pm$0.05 &    5 \\ 
587734891673026666   &    0.03061 &      9.84 &     9.79$\pm$0.06 &    1 \\ 
587732771584671859   &    0.03359 &      9.77 &     8.98$\pm$0.05 &    5 \\ 
587735349098971209   &    0.03254 &      9.77 &     8.4$\pm$0.11 &    5 \\ 
587735347486457882   &    0.03669 &     10.42 &     9.4$\pm$0.07 &    1 \\ 
587735431226130460   &    0.02706 &     10.56 &     8.84$\pm$0.06 &    5 \\ 
587735664773431424   &    0.03610 &     10.53 &     9.45$\pm$0.04 &    5 \\ 
587736584980463705   &    0.02300 &     10.29 &      9.36 &    3 \\ 
587736541491233170   &    0.02603 &      9.66 &     9.57$\pm$0.04 &    5 \\ 
587736808838594663   &    0.03669 &     11.27 &     9.99$\pm$0.05 &    1 \\ 
587738410325573774   &    0.02174 &      9.72 &     <8.58 &    5 \\ 
587738615415373880   &    0.03407 &     10.29 &     8.96$\pm$0.05 &    5 \\ 
587738067810189520   &    0.03639 &      9.55 &     <8.67 &    5 \\ 
587739130805420150   &    0.03479 &      9.54 &     9.33$\pm$0.04 &    5 \\ 
587739303684931731   &    0.03381 &      9.89 &     8.85$\pm$0.05 &    5 \\ 
587739504477077652   &    0.02593 &      9.64 &     9.07$\pm$0.04 &    5 \\ 
587739166775574594   &    0.03103 &     10.38 &     8.61$\pm$0.08 &    5 \\ 
587739609698140217   &    0.03276 &      9.61 &     8.24$\pm$0.09 &    5 \\ 
587739646208770144   &    0.02788 &      9.65 &     9.58$\pm$0.06 &    1 \\ 
587739828743962776   &    0.01578 &     10.69 &     10.11$\pm$0.17 &    1 \\ 
587739844856971464   &    0.03249 &      9.68 &     8.8$\pm$0.05 &    5 \\ 
587739845393186912   &    0.03396 &     10.47 &     10.01$\pm$0.06 &    1 \\ 
587741532777152653   &    0.02209 &      9.57 &     9.15$\pm$0.07 &    1 \\ 
587741722823950491   &    0.02388 &      9.82 &     <8.52 &    5 \\ 
587741815710744668   &    0.01919 &     10.02 &     8.76$\pm$0.05 &    5 \\ 
587741721214386184   &    0.03811 &      9.98 &     9.91$\pm$0.06 &    1 \\ 
587742010042744920   &    0.03526 &     10.87 &     10.18$\pm$0.05 &    1 \\ 
587742189908983921   &    0.02267 &      9.58 &     9.38$\pm$0.06 &    1 \\ 
587742576459251895   &    0.03537 &      9.91 &     9.17$\pm$0.05 &    5 \\ 
587742589333340203   &    0.03272 &     10.48 &     8.63$\pm$0.05 &    5 \\ 
587742188827443225   &    0.03066 &     10.49 &     <8.65 &    5 \\ 
587742189363527850   &    0.03229 &     10.19 &     <8.65 &    5 \\ 
587742551762796902   &    0.03291 &      9.58 &     9.97$\pm$0.05 &    1 \\ 
587742616170266870   &    0.03141 &      9.87 &     8.95$\pm$0.05 &    5 \\ 
587742627998531885   &    0.03511 &     10.29 &     9.71$\pm$0.06 &    1 \\ 
587742628523475193   &    0.03790 &     10.88 &     9.62$\pm$0.07 &    1 \\ 

& & & & \\
\hline
\end{tabular}
\tablecomments{Sources for the HI data given in the final column are 1=ALFALFA, 2=xGASS, 3=HI-MaNGA, 4=FASHI, 5= FAST (this work).}
\label{mhi_tab}
  \end{center}
\end{table}

\begin{table}
\addtocounter{table}{-1}
  \begin{center}
  \caption{HI gas masses for the PSB sample (continued).}
\begin{tabular}{lcccc}
\hline

SDSS ObjID & $z_{SDSS}$ & log M$_{\star}$ & log \mhi & HI source \\ 
           &   & (M$_{\odot}$) & (M$_{\odot}$) & \\
\hline
 & & & & \\
587742644626260055   &    0.03438 &     10.12 &     <8.57 &    5 \\ 
587745244159082657   &    0.02735 &      9.68 &     8.89$\pm$0.05 &    5 \\ 
587745403073855572   &    0.02970 &     10.14 &     <8.53 &    5 \\ 
588013382740279420   &    0.03417 &      9.85 &     <8.68 &    5 \\ 
588015508189872263   &    0.03253 &      9.52 &     9.88$\pm$0.05 &    5 \\ 
588015508208156794   &    0.01618 &      9.66 &     <8.53 &    5 \\ 
588016891172618376   &    0.02705 &     10.56 &     9.89$\pm$0.05 &    1 \\ 
588016839633666164   &    0.02230 &      9.90 &     8.52$\pm$0.06 &    5 \\ 
588017627780087861   &    0.02906 &     10.21 &     9.42$\pm$0.05 &    5 \\ 
588017724947759115   &    0.02408 &      9.83 &     9.58$\pm$0.05 &    1 \\ 
588848898848784582   &    0.03760 &      9.73 &       9.5 &    3 \\ 
588017992299380883   &    0.02310 &      9.59 &     9.64$\pm$0.05 &    1 \\ 
588023046941245549   &    0.02598 &      9.99 &     9.58$\pm$0.05 &    1 \\ 
588298664112881826   &    0.02911 &      9.63 &     9.34$\pm$0.04 &    5 \\ 

& & & & \\
\hline
\end{tabular}
  \end{center}
\end{table}

\smallskip

In our analysis of the HI gas fraction of PSBs in the following section, we will make use of the xGASS survey as a comparison sample.  Our FAST observations were therefore designed to match the sensitivity of the original GASS survey (Catinella et al. 2010), i.e. an HI gas fraction limit (\fgas\ = M$_{HI}$/M$_{\star}$) of 1.5 percent when \logm\ $>$ 10.5, and a HI gas mass \mhi\ =  $10^{8.7}$ \msun\ at lower stellar masses\footnote{The low mass extension of xGASS has a slightly modified detection threshold, but this does not affect the results presented here since we ultimately define our own effective detection thresholds that optimize both the xGASS sample and our own observations.}.  Throughout the analysis presented below (and in Tables \ref{tab:obs} and \ref{mhi_tab}) we adopt total stellar masses from the MPA/JHU catalog (see Kauffmann et al. 2003a).

\smallskip

The FAST observations were conducted in OnOff mode between September 2023 and May 2024. The central beam (M01) is used to switch between our targets (`source ON') and the blank sky (`source OFF'), with each `source ON' mode lasting for at most 6 minutes. The `source OFF' positions are selected to be $20'$ away from the targets, ensuring no contamination within $5'$. At the start of each ON/OFF cycle, a 10 K noise diode is turned on for 2 s for calibration. Data are recorded using the Spec(W+N) backend with a sampling time of 1 s. The backend has a bandwidth of 500 MHz and a channel width of 7.63 kHz, corresponding to 1.61 $\rm km~s^{-1}$ for HI observations at $z=0$.  

\smallskip

We extract slices of $\pm 2000~{\rm km~s^{-1}}$ around the optical velocities and reduce that portion of the spectrum. Radio frequency interference (RFI) is flagged manually in the waterfall map and interpolated using neighbouring data. Next, we mask channels that are $>3\sigma$ outliers for each sampling, subtract the baseline with a piecewise first-order polynomial function, and remove the standing wave with a sine function. The data are calibrated using the 10 K noise diode and converted to flux density with a zenith angle-dependent gain (Jiang et al. 2020). Spectra from each ON/OFF cycle and polarization are stacked and weighted according to the noise level. The spectrum is then rebinned to 20 \kms\ channels for the remaining analysis.

\smallskip

Figure \ref{atlas_fig} presents images of the first 12 galaxies in Table \ref{tab:obs} observed with FAST and their 21cm emission line spectra.  The remaining targets are shown in the Appendix  (Figure \ref{atlas_all}).  We modify the procedure of Yu et al. (2020) to find semi-connected (distance $<$50 /kms) segments of continuous positive channels within $\pm 500~{\rm km~s^{-1}}$ of the optical velocity for source detection. We set a $5\sigma$ threshold for each detected segment in the source-finding procedure to exclude possible false detections, which is validated by randomly sampling blank regions of the spectra. Among the 44 PSBs observed, 11 are non-detections.   For the 33 detected sources, the global HI flux is measured using the curve-of-growth method (Yu et al. 2020). HI masses are calculated by assuming a redshift-dependent distance:

\begin{figure*}
	\includegraphics[width=8.5cm]{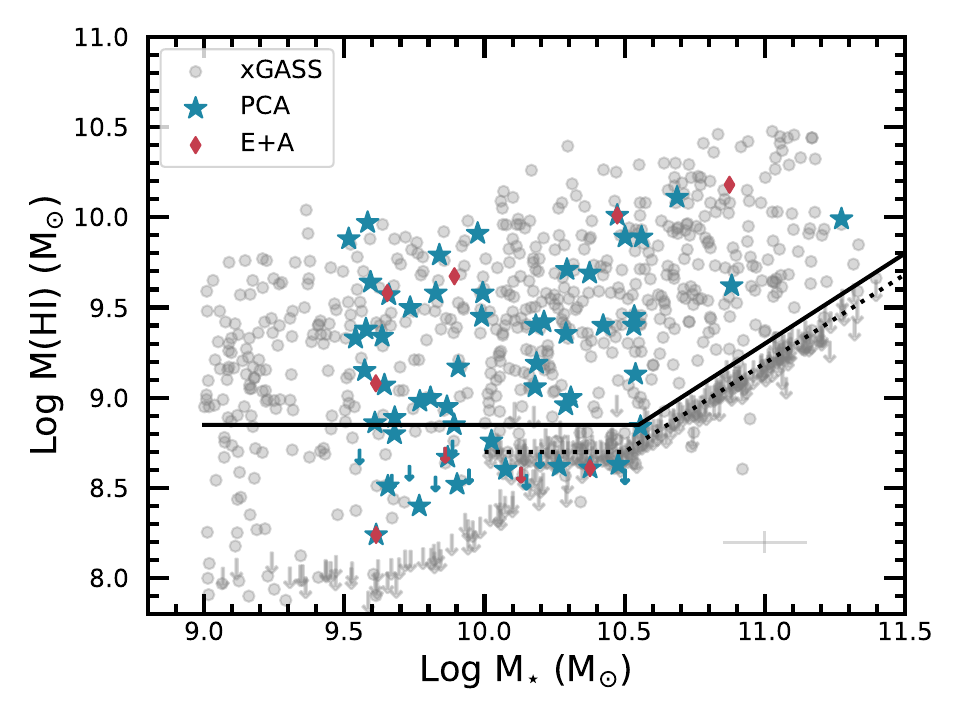}
	\includegraphics[width=8.5cm]{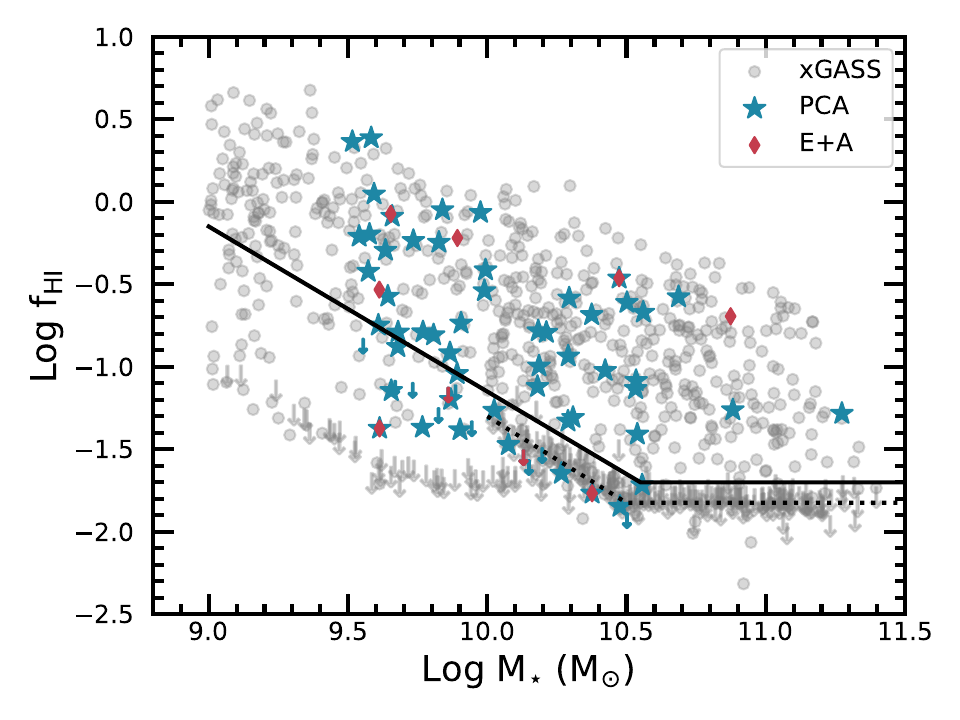}
        \caption{The HI mass (left) and the HI gas fraction (right) as a function of stellar mass for our PSB sample (coloured points) compared with the full xGASS (grey) sample.  The error bar in the lower right corner of the left hand panel shows the typical uncertainties in stellar mass and \mhi.  Upper limits (non-detections of HI) for both samples are shown with downward pointing arrows.  The dotted black line shows the GASS detection threshold.  The solid black line shows the slightly modified detection threshold adopted for the current work. }
        \label{fgas}
\end{figure*}

\begin{equation}
    M_{\rm HI} = \frac{2.356\times 10^5}{1+z}\left(\frac{d_{\rm L}(z)}{{\rm Mpc}}\right)^2 \left(\frac{\int F_v{\rm d}v}{\rm Jy~km~s^{-1}}\right) ~M_\odot,
\end{equation}

\noindent where $d_{\rm L}$ is the luminosity distance (using a cosmology of $H_0$ = 70 \kms\ Mpc$^{-1}$, $\Omega_M = 0.3$ and $\Omega_{\Lambda}$=0.7) and $\int F_v{\rm d}v$ is the integrated HI flux.  For the HI non-detections in galaxies more massive than \logm\ = 10.2, we compute 5$\sigma$ upper limits following the methodology adopted in the GASS survey, as described in Catinella et al. (2010), i.e. for a spectrum smoothed to 150 \kms\ and with an assumed 21 cm line width ($W_{21}$) of 300 \kms.  For galaxies less massive than \logm\ = 10.2, we follow the procedure for the low mass extension to GASS (xGASS, Catinella et al. 2018), smoothing to 100 \kms\ and assuming a line width of 200 \kms.  Thus the flux upper limit for each non-detection is then computed as

\begin{equation}
\int F_v{\rm d}v = 5 \times rms \times \sqrt{W_{21} \times dV} \times \sqrt{2}
\end{equation}

\noindent where $dV$ is the original 20 \kms\ channel width, the $rms$ for each non-detection is quoted in Table \ref{tab:obs}.  The $\sqrt{2}$ factor accounts for the additional smoothing, which for both stellar mass regimes is half of the assumed line width (i.e. 150 \kms\ and 100 \kms\ for galaxies with \logm\ above/below 10.2, respectively).

\smallskip

At the conclusion of our FAST observing campaign we have therefore obtained new observations for 44 targets, combined with 26 archival measurements, for a total of 70 galaxies.  However, two of these galaxies are dropped from our final sample. First, SDSS ObjID 587731681190740132 (observed as part of our FAST program)  had very strong RFI that precluded a reliable 21cm line flux measurement.  This target is listed in Table \ref{tab:obs}, but we do not calculate an HI mass.  Second, SDSS ObjID 587742567860469784, for which an archival HI measurement is available, was found to have a close companion that had not been identified in the original contamination check.  The final sample used in the analysis presented here therefore includes FAST HI masses or upper limits for 43 galaxies, plus 25 archival measurements for a total of 68 PSBs.  In Table \ref{mhi_tab} we present the HI masses for this sample, indicating the source of the HI data in the final column.

\smallskip

\section{Results}

\subsection{Comparison with xGASS}

In order to assess whether PSBs have a statistically distinct atomic gas fraction, a comparison sample is required.  For this purpose we make use of the xGASS survey (Catinella et al. 2018), which measured HI masses for a representative sample of $\sim$1200 galaxies in the nearby ($0.01<z<0.05$) Universe with stellar masses 9 $<$ \logm\ $<$ 11.5.  In both redshift and stellar mass, xGASS is therefore an excellent comparator for our PSB sample.

\smallskip

As described in the previous section, our FAST observations are designed to reach the same detection threshold in \mhi\ as xGASS.  In Figure \ref{fgas} we plot the HI masses (left panel) and gas fractions (\fgas\ = \mhi/M$_{\star}$, right panel) for the PSBs (coloured points) and xGASS (grey points) as a function of stellar mass.  The error bar in the lower right corner of the left hand panel shows the typical uncertainties in stellar mass and \mhi.  Different symbols are used for PSBs selected by either the traditional E+A technique (i.e. the Goto sample) or the PCA method (i.e. in the catalog of Wild et al. 2007).  Some PSBs appear in both catalogs, so the symbols are overlaid.  For both samples (PSBs and xGASS), upper limits (non-detections in HI) are shown as downward pointing arrows. The dotted black line shows the xGASS detection threshold, i.e. a gas fraction of 1.5 percent when \logm\ $>$ 10.5, and a HI gas mass limit of $10^{8.8}$ \msun\ at lower stellar masses.  However, in order to be more conservative, and to accommodate a handful of our PSB points with slightly shallower upper limits, for the purposes of this paper we adopt a modified version of the detection threshold.  The solid black line in Figure \ref{fgas} shows the modified detection threshold, which is a gas fraction of 2 percent when \logm\ $>$ 10.55, and a HI gas mass limit of $10^{8.85}$ \msun\ at lower stellar masses. It can be seen that, not only do the majority of xGASS non-detections sit comfortably below this slightly modified detection threshold, but all of our PSB non-detections do as well.

\smallskip

Figure \ref{fgas} visually illustrates one of the primary conclusions of this paper, that a high fraction of PSBs have readily detectable HI reservoirs.  Indeed, out of the sample of 68 galaxies, 57 (84 percent) have measured HI masses, with non-detections in only 11 cases.   Although the PSB sample is dominated by PCA selected galaxies (as expected, given the more inclusive selection criteria), both PCA and E+A PSBs have significant HI reservoirs.  Our conclusion that PSBs contain plentiful HI is therefore not dependent on the PSB selection method.  Figure \ref{fgas} also shows that, qualitatively, the PSBs sample a broad range of HI masses (spanning almost 2 dex) and gas fractions at a given stellar mass, including some cases that are as HI rich as the most gas rich xGASS galaxies.  The importance of a deep detection threshold compared with previous PSB studies is also highlighted by Figure \ref{fgas}.  For example, Chang et al. (2001) and Buyle et al. (2006) achieved HI mass limits of $\sim 3 \times 10^9$ \msun, which would have resulted in non-detections both for the majority of our PSBs, but also many of the normal galaxies in xGASS which will serve as our control sample. 

\smallskip

Whilst Figure \ref{fgas} provides a good visual demonstration that PSBs retain (at least some) of their HI reservoirs, in the following sections we move on to a more quantitative comparison with the xGASS sample.

\subsection{Detection fractions}

\begin{figure}
	\includegraphics[width=8.8cm]{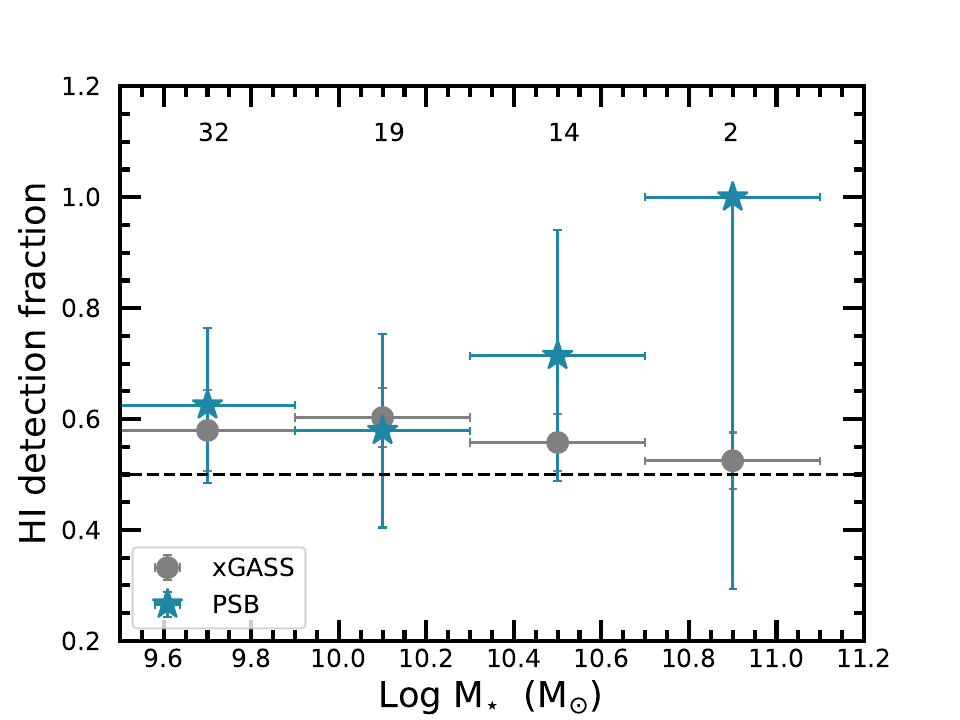}
        \caption{The HI detection fraction as a function of stellar mass for PSBs (blue stars) and xGASS (grey circles).  The error bars regions show the 1$\sigma$ errors based on binomial statistics (the positive y-error bars have been capped at a value of 1, since a fraction exceeding unity is impossible). The horizontal dashed line shows a detection fraction of 50 percent and the numbers along the top of the figure indicate the number of PSBs in each stellar mass bin. }
        \label{det_lim}
\end{figure}

A robust quantitative comparison between the HI gas fractions of the PSB sample with xGASS requires careful consideration of the inclusion of non-detections.  We begin in Figure \ref{det_lim} by quantifying the HI detection fraction of both the PSB and xGASS samples (due to the limited number of E+A PSBs, for the purposes of this assessment, we do not distinguish E+A and PCA PSBs).  The modified detection threshold adopted in this work (a gas fraction of 2 percent when \logm\ $>$ 10.55, and a HI gas mass limit of $10^{8.85}$ \msun\ at lower stellar masses, as shown by the solid black line in Figure \ref{fgas}) is used to assess whether any given data point is considered a detection or an upper limit.  If the galaxy falls below the solid black line in Figure \ref{fgas}, regardless of whether it has actually been detected or not, it is considered a non-detection.  If a galaxy has an HI detection above the black line, it is considered a detection.  Finally, if a galaxy is a non-detection above the black line, it is insufficiently deep to make a detection assessment and is discarded from the sample (this happens in only a handful of cases in xGASS, and never in our PSB sample).  Figure \ref{det_lim} shows the detection fractions for the PSBs (blue points) and xGASS (grey points) as a function of stellar mass, in bins of width 0.4 dex (starting at \logm\ = 9.5).  The error bars show the 1$\sigma$ uncertainty based on binomial statistics.  The horizontal dashed line shows a detection fraction of 50 percent and the numbers along the top of the figure indicate the number of PSBs in each stellar mass bin.  Since one PSB has a stellar mass above the upper bin limit, these numbers only add up to 67, rather than the 68 that are in the full sample.  

\smallskip

Two important conclusions can be drawn from Figure \ref{det_lim}.  First, the figure shows that the HI detection fraction is greater than 50 percent for both samples across the full stellar mass range for the mass bins we have chosen (with the caveat that there are only two galaxies in the highest mass bin, so statistics above \logm\ $>$ 10.7 are poor).  This means that we will later be able to compute the median HI gas fraction in these stellar mass bins, even though there are non-detections in the sample.  Second, we find that the detection fraction of HI in the PSBs is either consistent with (for \logm\ $<$ 10.3) or possibly even greater than (for \logm\ $>$ 10.3) that of the xGASS sample, although with the large uncertainties, the difference is not statistically significant.  Nonetheless, Figure \ref{det_lim} provides complementary support of our conclusion from Figure \ref{fgas} that PSBs are not systematically depleted of HI compared with the xGASS sample at fixed stellar mass.

\subsection{Gas fractions}

\begin{figure*}
	\includegraphics[width=8.5cm]{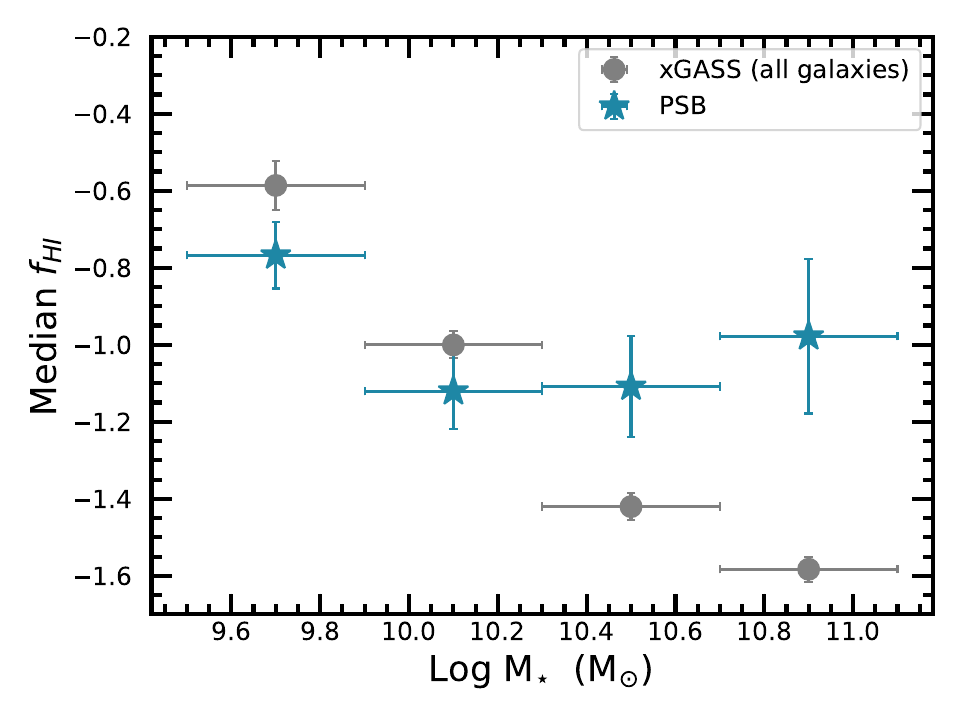}
	\includegraphics[width=8.5cm]{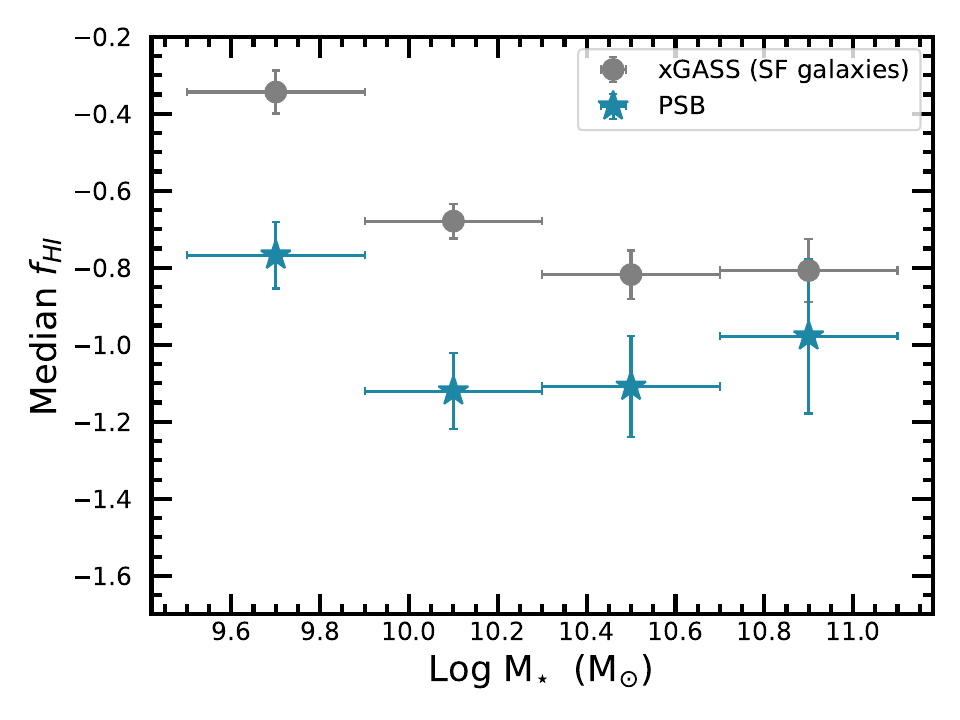}
        \caption{The median HI gas fraction for PSBs (blue stars) and xGASS (grey circles) as a function of stellar mass. The blue points are the same in both panels, but different populations of xGASS are shown in each panel. In the left panel the xGASS median is computed for all galaxies.  In the right panel the xGASS median is computed only for star-forming galaxies.  The figure shows that, at fixed $M_{\star}$ and compared with all xGASS galaxies, PSBs are either marginally HI-poor (\logm\ $<$ 10.3) or significantly HI-rich (\logm\ $>$ 10.3).  However, if considered relative to their (presumed) progenitor population of star-forming galaxies, PSBs are relatively HI-poor. }
        \label{med_fgas}
\end{figure*}

Having established that the detection fraction is at least 50 percent in both galaxy samples (xGASS and PSBs) we can now make a quantitative comparison of the HI gas fractions.  In the left panel of Figure \ref{med_fgas} we plot the median gas fraction, as a function of stellar mass, for the PSB and xGASS samples.  As previously asserted, these medians are robust, despite the presence of upper limits, because the detection fraction is above 50 percent for these bins (i.e. upper limits provide no information in the calculation of a median because they represent less than half of the sample).  The left panel of Figure \ref{med_fgas} shows that the PSBs have a marginally lower gas fraction, by $\sim$0.1 - 0.2 dex, at the lower stellar mass end of the sample  (\logm\ $<$ 10.3) but the statistical significance is not high.  Conversely, at \logm\ $>$ 10.3 the gas fraction of PSBs now exceeds the comparison xGASS sample at the same stellar mass by 0.4 - 0.6 dex at the highest stellar masses (although we remind the reader that our statistics above \logm\ $>$ 10.7 are poor).  This higher gas fraction is consistent with the higher detection fraction amongst PSBs compared with xGASS at these stellar masses shown in Figure \ref{det_lim}.   

\smallskip

The results in the left panel of Figure \ref{med_fgas} suggest the potentially surprising result that, far from being depleted of their neutral atomic gas, PSBs are enriched with HI at high stellar masses.   However, it is important to appreciate that the nature of the galaxy population (in general) evolves strongly with stellar mass.  Below \logm\ $\sim 10$, the majority of galaxies are star-forming.  However, as the stellar mass increases, an ever growing fraction of galaxies becomes passive, as has been shown many times in large galaxy surveys.  The relevance of the evolving population demographics as a function of stellar mass is that we expect the progenitors of PSBs to be star-forming galaxies.  Therefore, the germane comparison sample is perhaps \textit{not} between PSBs and all xGASS galaxies at the same stellar mass but only the star-forming population.

\smallskip

We therefore repeat our comparison of median gas fractions between xGASS and PSBs, but now only include star-forming galaxies (as selected by using the Kauffmann et al. 2003b cut in [OIII]/H$\beta$ -- [NII]/H$\alpha$ space) for the xGASS sample.  In contrast to the results in the left panel of Figure \ref{med_fgas}, which used all galaxies in xGASS, the right-hand panel of the figure shows that PSBs are $\sim$ 0.2--0.4 dex less HI-rich than their comparison sample in all stellar mass bins.  Therefore, under the assumption that the progenitors of PSBs are star-forming galaxies, the right panel of Figure \ref{med_fgas} suggests that, despite retaining abundant HI gas, the PSBs have indeed lost/consumed some of their atomic gas reservoirs in the quenching process.

\smallskip

It would be potentially interesting to also compare the median gas fraction of PSBs to quenched galaxies in xGASS, i.e. the population into which they are evolving.  However, the low detection fraction of HI in quiescent galaxies means that robust median gas fractions can not be computed for xGASS.  We can nonetheless make a more generalized comparison of our results to studies of HI in elliptical galaxies in the nearby Universe.  In what remains one of the definitive works on this topic, Serra et al. (2012) measured the HI mass for 166 nearby early type galaxies in the ATLAS$^{3D}$ survey, down to sensitivity thresholds two orders of magnitude deeper than xGASS (and our PSBs).  Although some of the (field) ellipticals were found to be as HI-rich as spirals, Serra et al. (2012) found that the vast majority of their sample had HI masses below our detection threshold of \logm\ $\sim$ 9.0. Likewise, whereas we find median PSB gas fractions of $\sim0.1$ (Figure \ref{med_fgas}), most ellipticals have gas fractions 10 times lower than this (Serra et al. 2012).  In summary, PSBs have HI gas masses (and fractions) intermediate between star-forming galaxies and quiescent ellipticals. 

\subsection{Gas mass offsets}\label{dmhi_sec}

Figures \ref{det_lim} and \ref{med_fgas} demonstrate that, as a population, PSBs remain endowed with significant reservoirs of HI gas.  However, these previous figures do not allow us to quantify the HI deficit or excess on a galaxy-by-galaxy basis, which would, in turn, allow us to assess the variation in the population, as well as look for trends in HI-richness as a function of galaxy property.  We therefore compute, for each PSB galaxy, an HI mass offset, $\Delta$\mhi, which is designed to capture the relative HI excess or deficit of an individual galaxy compared to xGASS galaxies of the same stellar mass.  Whilst $\Delta$\mhi\ is similar in spirit to the well-known HI deficiency paramater (Haynes \& Giovanelli 1984), our metric is computed at fixed stellar mass, rather than for a given morphological type and size\footnote{Additionally, HI deficiency is positive for gas-poor galaxies, whereas our metric uses the, we believe, more intuitive convention that a positive value indicates a gas rich galaxy.}.

In order to compute $\Delta$M(HI), we assemble a stellar mass matched control sample from the xGASS survey, selecting only galaxies whose M$_{\star}$ is within 0.1 dex of the PSB\footnote{Our calculation of $\Delta$\mhi\ is similar to the measure of HI richness proposed by Guo et al. (2021), in which the offset of a galaxy is computed from a parametric fit of \mhi\ vs. M$_{\star}$ for star-forming galaxies.  The main difference between the two approaches is that rather than fitting the \mhi\ main sequence, we sample it for a given stellar mass.  The HI richness computed using the Guo et al. (2021) formula results in $\Delta$\mhi\ values $\sim$ 0.1 dex lower than our calculation. Our qualitative conclusion about the persistance of a significant HI reservoir in PSBs is therefore unaffected.}.  Typically 50--150 xGASS galaxies are matched to each PSB.  $\Delta$\mhi\ is then computed as the difference (in log space) between the HI mass of the PSB galaxy and the median of the control sample.  In this way, a $\Delta$\mhi\ = 0 indicates a PSB galaxy that has the same HI mass as expected for its stellar mass, according to the comparison galaxies in xGASS. $\Delta$\mhi\ can be computed for all but the highest stellar mass PSB in our sample (objID= 587736808838594663 with \logm\ = 11.27).  For this one high mass PSB, with the relatively narrow mass matching tolerance that we have imposed, the xGASS detection fraction is less than 50 percent and the median value is hence not accurately measured, resulting in a lower limit $\Delta$\mhi $>$0.65 dex (i.e. still substantially elevated).  

\begin{figure*}
	\includegraphics[width=8.5cm]{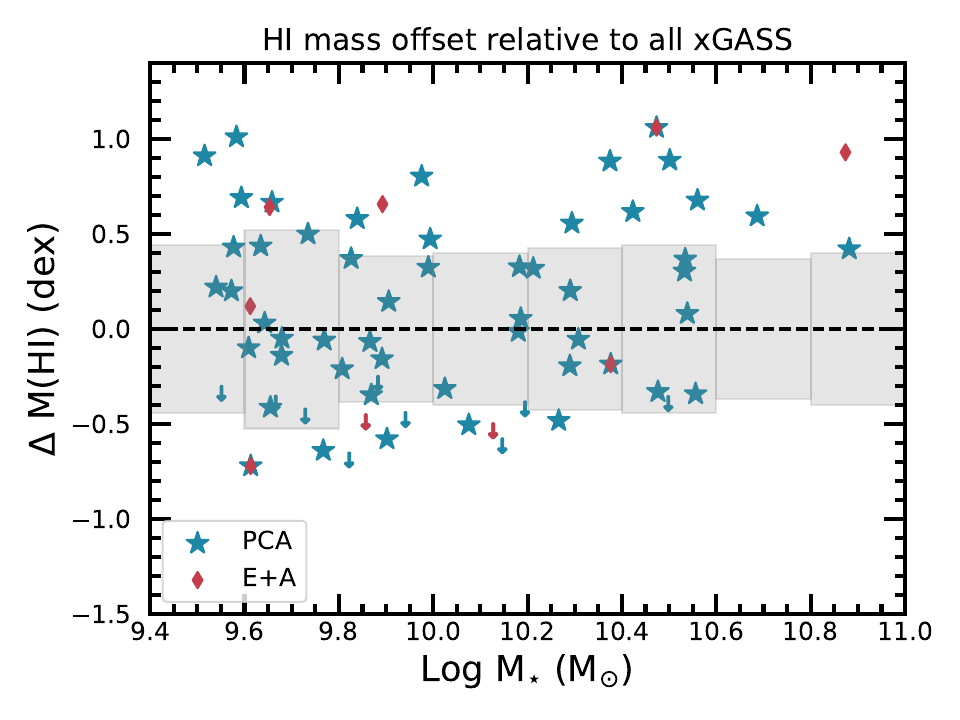}
	\includegraphics[width=8.5cm]{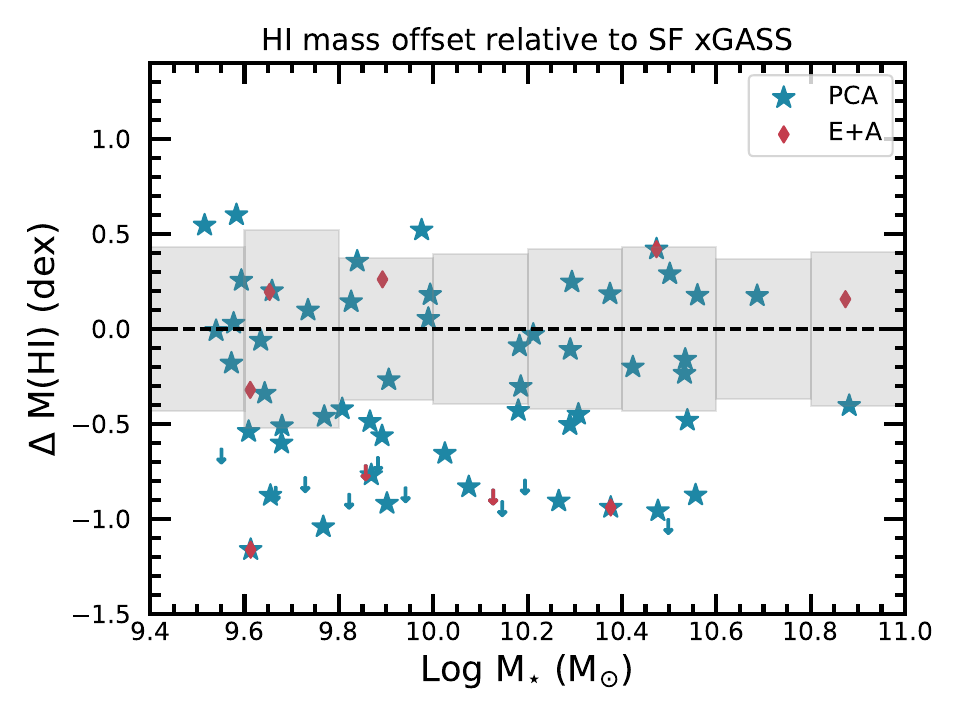}
        \caption{The HI gas mass offset of PSBs compared with a stellar mass matched control sample drawn from xGASS.  In the left hand panel, the control galaxies are drawn from the full xGASS sample.  In the right hand panel only star-forming control galaxies are used.  The dashed line shows a value of $\Delta$\mhi\ = 0, i.e. an \mhi\ that is statistically consistent with the control sample. The grey shaded area shows the $\pm 1 \sigma$ range of $\Delta$\mhi\ for galaxies in xGASS (all, or star-forming for the left and right hand panels respectively) in bins of width $\pm$0.2 dex.  The figure shows that PSBs have a diversity of HI gas mass offsets, ranging from being even more gas rich than star-forming galaxies to a factor 10 more gas-poor.}
        \label{delta_fgas}
\end{figure*}

\smallskip

The left panel of Figure \ref{delta_fgas} shows the distribution of $\Delta$\mhi, once again distinguishing between those that have been selected by either the E+A or PCA (or both) method.  The horizontal dashed line shows a value of $\Delta$\mhi\ = 0, i.e. gas-normal for its stellar mass.  The grey shaded area shows the $\pm 1 \sigma$ range of $\Delta$\mhi\ for galaxies in xGASS in bins of width $\pm$0.2 dex in order to demonstrate the range in the underlying population (i.e. the scatter in the atomic gas main sequence).  Since this scatter dominates over the uncertainty in the measurement of \mhi\ for any individual galaxy, we do not assign error bars to our $\Delta$\mhi\ values.  Moreover, since the scatter for xGASS can't be meaningfully calculated at \logm\ $>11$, we do not extend the plot to cover the highest stellar mass PSB (\logm\ = 11.27).  However, we iterate that this PSB has a lower limit of $\Delta$\mhi\ $>$ 0.65 dex, making it very HI rich for its stellar mass.

\smallskip

The distribution of $\Delta$\mhi\ amongst PSBs shown in the left hand panel of Figure \ref{delta_fgas} now allows us to more clearly see the distribution of gas properties compared with a control sample at fixed stellar mass.  A handful of PSBs are relatively gas poor, containing only $\sim$ one third the amount of HI gas expected for their stellar mass (i.e. $\Delta$\mhi\ = $-0.5$ dex).  However, the majority of PSBs are HI-normal for their stellar mass (i.e. they lie within the grey shaded region), or gas rich by up to a factor of 10.  This is true for both E+A and PCA selected PSBs.  

\smallskip

As discussed in the previous sub-section, our conclusions regarding the HI properties of PSBs can change dramatically depending on the choice of comparison sample (i.e. all galaxies or only star-forming ones).  Therefore we re-compute $\Delta$\mhi, but now using only star-forming galaxies in xGASS for the control sample.  The right-hand panel of Figure \ref{delta_fgas} plots the re-calculated $\Delta$\mhi\ on the same axis scale as the right-hand panel.  Once again, the grey box shows the $\pm 1 \sigma$ spread of $\Delta$\mhi\ for the star-forming galaxies in xGASS (which is marginally smaller than the scatter amongst the full sample).  As expected, using only the star-forming galaxies in the control sample leads to lower overall $\Delta$\mhi\ values in the PSBs.  Nonetheless, approximately half of the PSBs have $\Delta$\mhi\ within the 1$\sigma$ scatter of star-forming galaxies in xGASS, with the remainder more gas-poor.  In summary, when considered as the descendants of star-forming galaxies (right panel of Figure \ref{delta_fgas}), PSBs are either `gas-typical' or gas-poor by factors of $\sim$ 3--10.

\subsection{PSBs in the context of the green valley}\label{gv}

In the previous subsections we have demonstrated that, although a high fraction of PSBs host detectable amounts of atomic gas, they have HI gas fractions that are nonetheless either typical of, or somewhat depleted compared to their presumed progenitor population of star-forming galaxies.  In this subsection, we next compare the PSBs, which represent the very particular case of recent and rapid quenching, to the general green valley population that are also believed to be in transition from star-forming to quiescent.

\smallskip

The relatively sparse population of the green valley (defined either through intermediate specific SFRs or colours, e.g. Wyder et al. 2007; Brinchmann et al. 2004; Salim et al. 2007; Cortese \& Hughes 2009; Salim 2014) was once thought to indicate that galaxies must transition rapidly from star-forming to passive.  However, based on both observations (e.g. Schawinski et al. 2014; Peng et al. 2015; Trussler et al. 2020; Tacchella et al. 2022) and simulations (Rodriguez-Montero et al. 2019; Walters, Woo \& Ellison 2022) it is now understood that galaxies take diverse transitory routes, and can linger in the green valley for many Gyr.  It is therefore of interest to compare the gas fractions of PSBs (fast quenchers) to the general green valley population (diverse quenching timescales), in order to assess whether their recent, rapid quenching sets them apart.

\smallskip

The most natural way to place PSBs in the context of the general quenching population is probably to compare them with green valley galaxies that exhibit the same middling SFRs.  However, determining SFRs for PSB galaxies is notoriously uncertain for several reasons.  First, PSBs are deliberately selected to have weak emission lines, undermining the usual adoption of H$\alpha$ as a robust SFR indicator.   For those PSBs with detectable H$\alpha$, AGN contamination is often an issue (e.g. Yesuf et al. 2014; Pawlik et al. 2018; Ellison et al. 2022).  Finally, the use of D4000 (commonly used to circumvent these issues, e.g. Brinchmann et al. 2004) is compromised since it is sensitive to longer timescales than H$\alpha$ and is therefore likely to be more indicative of the pre-quenching SFR than the current one.  Although it is possible to obtain SFRs for PSBs through full spectral fitting (which we will indeed carry out in the next sub-section), these fits are sufficiently computationally expensive that it is impractical to perform them for a large reference sample (e.g. all of xGASS).

\smallskip

We therefore turn to $NUV-r$ colour as an alternative indicator of sSFR (e.g. Salim 2014), since the $NUV$ and $r$ bands are proxies for current and past-integrated star formation, respectively.  The exact definition of the green valley in $NUV-r$ space varies between works, but is typically in the range of $4 < NUV-r < 5$ (e.g. Salim 2014) to $3 < NUV-r < 5$ (e.g. Brown et al. 2015), with the possibility of a stellar mass dependence (e.g. Coenda et al. 2018).  In the results presented here, we adopt a working definition of the green valley to be  $3 < NUV-r < 5$, noting that our qualitative conclusions are not sensitive to the precise value of the boundaries.  Indeed, regardless of the definition, Catinella et al. (2018) have shown that the xGASS sample exhibits a tight anti-correlation between \fgas\ and $NUV-r$, that holds across both the star-forming and green valley populations (although the relation is poorly constrained for redder colours due to the low detection fraction in xGASS).  A similar trend between gas fraction and colour is also seen for redder band passes, e.g. in $NUV-H$ (Cortese \& Hughes 2009).

\smallskip

\begin{figure}
	\includegraphics[width=8.5cm]{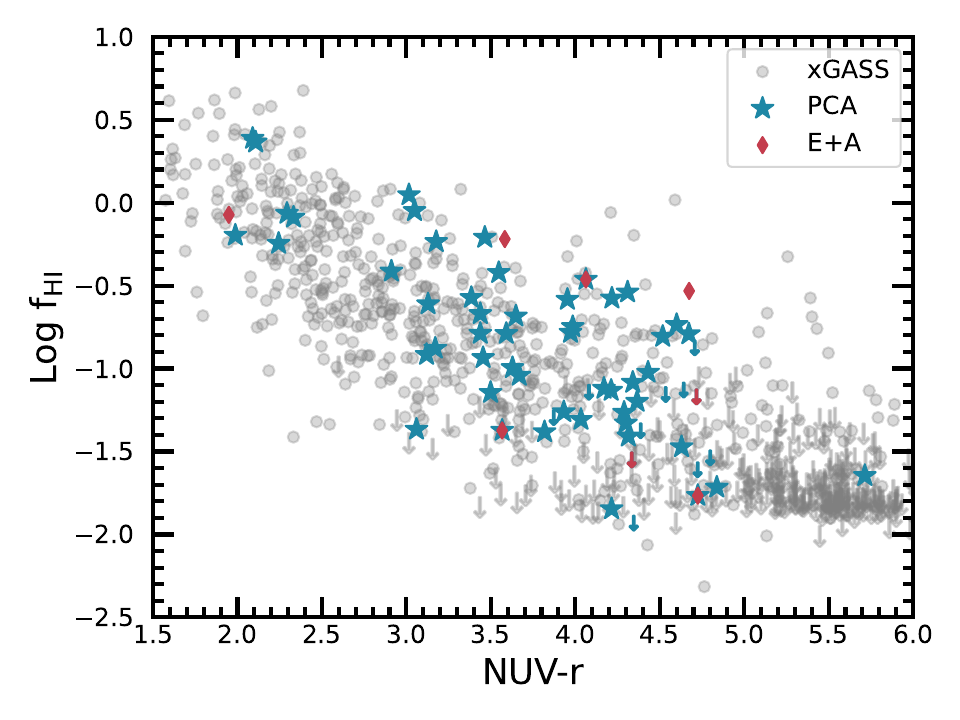}
        \caption{The relation between \fgas\ and $NUV-r$ (a proxy for sSFR) for xGASS (grey points) and PSBs (coloured points).  The green valley is typically considered to have colours in the range $3<NUV-r<5$.  Most PSBs are consistent with this range of green valley colours, although a small number are still as blue as actively star forming galaxies ($NUV-r<3$).  Regardless of $NUV-r$ colour, the gas fractions of PSBs fall broadly within the scatter of xGASS galaxies of the same colour.}
        \label{nuvr}
\end{figure}

In order to obtain $NUV-r$ colours for both our PSB sample and xGASS, we cross match our sample to the NASA Sloan Atlas (NSA\footnote{http://nsatlas.org/data}, version v1\_0\_1), adopting the $k$-corrected rest-frame magnitudes derived from elliptical Petrosian apertures.  Figure \ref{nuvr} shows the well-known anti-correlation for xGASS (grey points) with the PSBs overlaid (coloured points).  First of all, we note that the PSBs mostly lie, as expected, in the range of $NUV-r$ colours expected for the green valley ($3<NUV-r<5$), although a small number as blue as actively star forming galaxies ($NUV-r<3$).  A similar range of intermediate and blue colours for `central' PSBs (as we expect ours to be, given that they are fibre selected) was previously found by Cheng et al. (2024).  Figure \ref{nuvr} also shows that, regardless of $NUV-r$ colour, the gas fractions of PSBs fall broadly within the scatter of xGASS galaxies of the same colour, although the visual impression is that they are somewhat skewed towards higher gas fractions.

\smallskip

\begin{figure}
	\includegraphics[width=8.5cm]{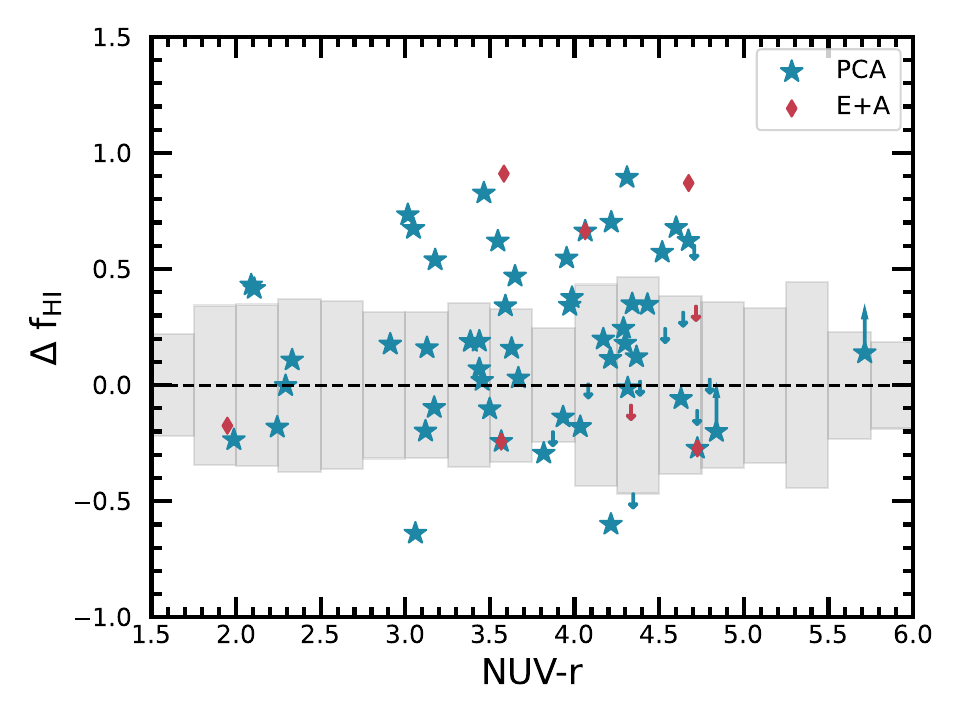}
        \caption{Gas fraction offset for PSBs compared with an $NUV-r$ matched control sample drawn from xGASS.     The dashed line shows a value of $\Delta f_{\rm HI}$ = 0, i.e. a gas fraction that is statistically consistent with the control sample. The grey shaded area shows the $\pm 1 \sigma$ range of $\Delta f_{\rm HI}$ for galaxies in xGASS as a function of $NUV-r$ in bins of width 0.25 magnitudes.  For the two reddest galaxies in our sample, less than 50 percent of the control sample have detections, so their $\Delta f_{\rm HI}$ is a lower limit and shown with an upwards arrow.  The figure shows that, compared with other green valley galaxies, PSBs have a diversity of gas fraction offsets, with two-thirds lying within the scatter of gas fractions exhibited by other green valley galaxies, and the remaining third with elevated values. }
        \label{dnuvr}
\end{figure}

In order to quantify the gas fractions of the PSBs relative to green valley galaxies, we once again compute an offset compared with a matched sub-sample of xGASS galaxies.  However, this time we calculate an HI gas fraction offset relative to xGASS galaxies with the same $NUV-r$ (as opposed to our original calculation of $\Delta$\mhi\ which was computed at fixed $M_{\star}$).  Therefore, for each PSB we compute a $\Delta$\fgas\ which is the difference (in log space) between the HI gas fraction of a given PSB and the median HI gas fraction of all xGASS galaxies whose $NUV-r$ matches the PSB to within $\pm$ 0.1.  $\Delta$\fgas\ is successfully computed for all but the two reddest PSBs in our sample.  For these remaining two the detection fraction in xGASS drops below 50 per cent such that the median of the matched comparison sample can not be robustly calculated. For these two cases we thus only have an upper limit for the median \fgas\ in the control sample, in turn leading to a lower limit for the PSB $\Delta$\fgas.  

\smallskip

In Figure \ref{dnuvr} we show the distribution of $\Delta$\fgas\ for the PSBs as a function of $NUV-r$.  The two reddest PSB galaxies, for which we only have a lower limit for $\Delta$\fgas, are plotted with upward arrows. The grey shaded regions shows the $\pm 1 \sigma$ range $\Delta$\fgas\ of xGASS galaxies (which can be computed in the same way as for the PSBs), computed in $NUV-r$ bins of width 0.25 magnitudes, and thus illustrate the underlying scatter of the \fgas\ versus $NUV-r$ relation shown in Figure \ref{nuvr}.  Approximately two thirds of the PSBs shown in Figure \ref{dnuvr} lie within the scatter of the xGASS comparison sample, illustrating that the majority of PSBs have `normal' gas fractions when compared to a population with similar specific SFRs.  Put another way, most PSBs have gas fractions typical of those measured in other green valley galaxies.  However, about one third of the population is significantly gas-rich, with atomic gas fractions three to ten times that of the typical green valley population.

\subsection{Gas fractions as a function of time since burst}

Several previous papers have found tentative evidence that both the molecular gas (French et al. 2018a; Bezanson et al. 2022) and dust (Smercina et al. 2018; Li et al. 2019) decline with time since burst.  These works paint a picture in which quenching is initiated despite the presence of a molecular gas reservoir, but that the state of the ISM is significantly affected in the aftermath of the starburst.  With our large sample of HI measurements we are now in a position to assess whether the atomic gas content evolves with post-starburst age.

\smallskip

In order to model the star formation histories of the galaxies in our sample we make use of the Bayesian Analysis of Galaxies for Physical Inference and Parameter EStimation (\textsc{bagpipes}) code (Carnall et al. 2018).  We follow the application of \textsc{bagpipes} to the SDSS spectra of our PSB sample as described in Leung et al. (2024).  Briefly, the spectral energy distribution is fit using a two-component star formation history, comprising an older exponentially declining population coupled with a burst.  Gaussian process noise is included in the model's predicted spectrum to help account for correlated observational uncertainties.  The model has 15 free parameters including the time since the peak of the starburst ($t_{burst}$).   The primary difference between our fits and the approach described in Leung et al. (2024) is that the Balmer emission lines are not masked out, but rather fit simultaneously with the stellar continuum.  49 of our PSBs are successfully fitted, 40 HI detections and 9 non-detections; failed fits include those where the extinction has reached the maximum set by the prior or with excessive Gaussian process noise indicating that the fit is unreliable.

\smallskip

\begin{figure}
	\includegraphics[width=8.5cm]{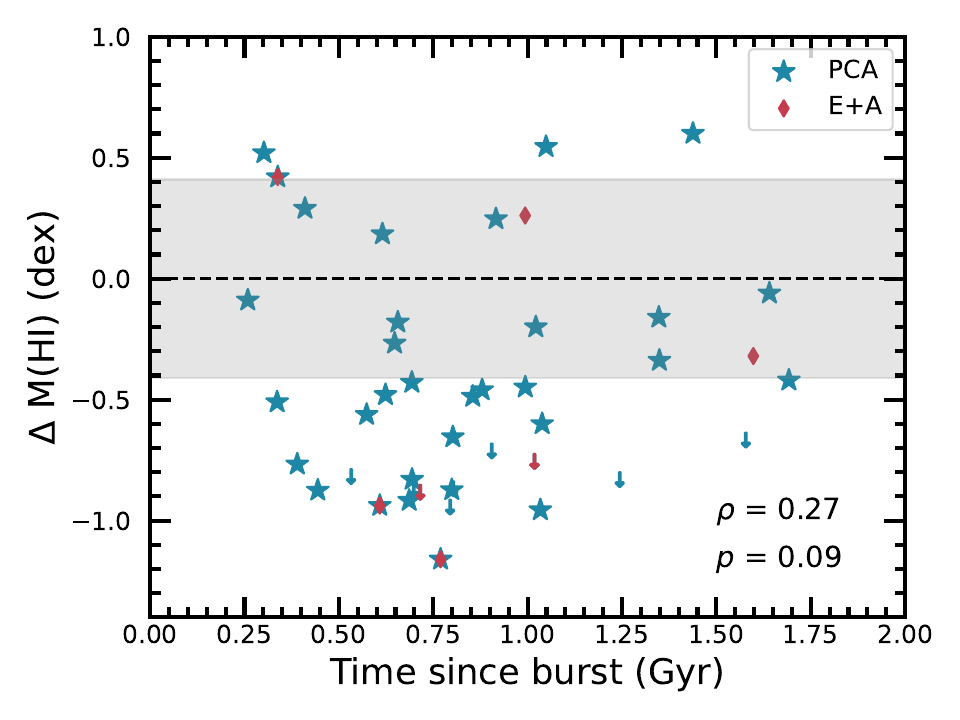}
        \caption{Atomic gas mass offset ($\Delta$\mhi), calculated relative to mass-matched samples of star-forming xGASS galaxies) as a function of time since burst for PSB galaxies with \logm\ $<$ 10.8 and successful \textsc{bagpipes} fits.  The grey shaded region shows the RMS scatter of $\Delta$\mhi\ in the full xGASS sample and thus demonstrates the range expected within the general galaxy population.  The Pearson correlation coefficients and probabilities are reported in the lower right of the figure.  There is no trend between atomic gas mass offset and time since burst, indicating that the global HI mass reservoir is not systematically affected in the aftermath of the starburst.}
        \label{bage}
\end{figure}

In Figure \ref{bage} we plot the HI gas mass offset ($\Delta$\mhi) versus the time since burst.  The $\Delta$\mhi\ values shown in Figure \ref{bage} are those shown in the right-hand panel of Figure \ref{delta_fgas}, in which the control sample are star-forming galaxies in xGASS.   Visually, there is no trend between the HI gas mass offset and $t_{burst}$.  Even PSBs that experienced their burst over 1 Gyr ago generally retain HI gas masses within the scatter expected from the atomic gas main sequence (shown in grey), with some significantly more HI-rich.  A quantitative assessment confirms the visual impression of a lack of correlation between $\Delta$\mhi\ and $t_{burst}$; a Pearson correlation test returns values of $\rho=0.27$ and $p=0.09$.  Likewise, there is no correlation if we use \mhi\ or \fgas\ instead of $\Delta$\mhi.  We also checked for correlations between $\Delta$\mhi\ and both burst fraction and decay time of the burst and similarly found no statistical signal.  We therefore conclude that, in contrast to previous tentative evidence in the literature for molecular gas (French et al. 2018a; Bezanson et al. 2022), there is no systematic depletion of the global HI reservoir as a function of time after the starburst.

\section{Discussion}

The over-arching conclusion of the work presented here is that recently and rapidly quenched post-starburst galaxies can retain significant atomic gas reservoirs.  Although PSBs are identified via very specific criteria, previous work has also demonstrated that the more general population of galaxies in transition between star-forming and quiescent exhibit a wide diversity of atomic gas fractions.  In one of the earliest statistical studies of the HI properties of transitioning galaxies, Cortese \& Hughes (2009) found both HI deficient and HI normal populations.  Whilst the former population was dominated by cluster galaxies, suggesting that their HI poverty and commensurate quenching was linked to stripping, HI normal transition galaxies were often found to be in mergers.  In this sense, there is qualitative similarity between our findings and those of Cortese \& Hughes (2009), because PSBs (found here to often be gas normal) are frequently mergers (e.g.  Pawlik et al. 2018; Sazonova et al. 2021; Wilkinson et al. 2022; Ellison et al. 2024).  In a more recent study Janowiecki et al. (2020) showed that green valley galaxies can be as HI rich as star-forming galaxies, consistent with the results in Figure \ref{delta_fgas} where approximately half of the PSB sample is found to have $\Delta$\mhi\ within the scatter of star-forming galaxies in xGASS.  Perhaps even more extreme is the recently discovered population of extremely HI-rich, fully quenched galaxies presented by Li et al. (2024).  Taken together, these works demonstrate that low SFRs and atomic gas poverty need not go hand-in-hand.

\smallskip

It could be argued that HI is irrelevent for the status of current star formation in galaxies.  For example, Bigiel et al. (2008) showed that star formation rate surface densities are uncorrelated with the surface density of HI and Catinella et al. (2018) showed that galaxies on the star forming main sequence have HI gas fractions spanning over 2 orders of magnitude.  Nonetheless, atomic hydrogen is the initial reservoir from which molecular gas is expected to form.  If star formation has ceased despite the presence of ample HI, it is therefore germane to ask why.  One possibility is that, whilst the mass of HI gas can be significant, it may be diffuse.   For example, Lemonias et al. (2014) mapped HI in a sample of 20 nearby galaxies with suppressed SFRs but high HI masses and found surface densities below the critical value for star formation identified by Bigiel et al. (2008).  However, these surface densities were considered on a galaxy-wide scale (within $R_{90}$) whereas star formation thresholds are usually assessed on sub-kpc scales.  In counterpoint to the idea of low gas surface densities leading to quenching, Bigiel et al. (2010a) found that low level star formation proceeds in the outer disk of M83, despite lower columns of HI than found in the main star-forming disk.  With a larger sample of galaxies, Bigiel et al. (2010b) confirm a correlation between far-UV luminosity (tracing massive star formation) and HI in outer disks, suggesting that HI is a critical component for setting star formation efficiencies in these low density environments.

\smallskip

Kpc-scale investigations of PSBs have, to our knowledge, only been performed by Klitsch et al. (2017), who mapped HI in two local PSB galaxies and found fairly typical distributions of atomic gas.  High turbulence, detected in the molecular gas phase of PSBs, has also been proposed as a bottleneck in the transformation of diffuse gas into stars (Otter et al. 2022; Smercina et al. 2022).  However, a re-assessment of some of these datsets has revised previous turbulent pressure measurements downwards, casting doubt on this as a mechanicm for quenching (Sun \& Egami 2022).  Moreover, Klitsch et al. (2017), in their HI observations of two PSBs, found regularly rotating HI.  High resolution (kpc-scale) mapping HI for a larger sample of post-starbursts, to determine both surface densities and model the atomic gas kinematics, is clearly an urgent direction for future study.

\smallskip

An alternative intriguing possibility for the presence of HI in PSBs is that it has been recently delivered by a merger.  In this scenario, the atomic gas measured in PSBs is not the long-lived remnant of the pre-burst galaxy, but rather a recent acquisition by a previously passive galaxy.  Cortese \& Hughes (2009) considered this scenario as an explanation for HI-normal galaxies in the green valley, noting a high merger fraction amongst this population.  Such galaxies may have had historically low SFRs, but just recently been the beneficiaries of fresh gas delivery. Indeed, Ellison et al. (2018) showed that summing the stellar mass and HI gas mass components of randomly selected pairs of galaxies (i.e. simulating the effect of a merger) leads to remnants that are offset above the normal atomic gas main sequence.  However, we consider this scenario an unlikely explanation for the presence of abundant HI in PSBs. In Ellison et al. (2022) we used \textsc{bagpipes} to fit the star formation histories of PSBs in SDSS and found the overwhelming majority had SFRs consistent with the main sequence 1-2 Gyr ago.  I.e. the progenitors of PSBs are star-forming, rather than quiescent galaxies that have been rejuvenated.  Moreover, we find that the mergers and non-mergers in our PSB sample both show a range of gas fractions, with no systematic difference between the two.  

\smallskip

Given their HI reservoirs, an intriguing possibility is whether or not PSBs may re-ignite their star formation at some future time.  The persistance of abundant HI even a Gyr after the burst certainly indicates that the atomic reservoir will be long lived.  Indeed, previous spatially resolved studies have found that centrally concentrated PSBs (as we expect ours to be, given that they are based on single fibre SDSS spectroscopy) have already fully quenched in their outskirts, with the inner regions the last to shut-down star formation (Li et al. 2023; Cheng et al. 2024).  I.e. central PSBs such as those in our sample are close to completing galaxy-wide quenching, despite the persistence of the HI found in our study.  We are currently undertaking an observing campaign to obtain molecular gas masses from CO(1-0) for the same sample for which we have presented HI herein (Rasmussen et al., in prep).  Combining the atomic and molecular gas masses for the same PSB sample will provide a unique insight into the interplay of the gas phases.  For example whether we see the molecular gas fractions declining with time since burst (which would be in agreement with previous studies) despite the long-lived HI.  Alternatively, with our larger sample and state-of-the-art star formation histories we may find that the molecular gas also remains constant.

\smallskip

Finally, our analysis has highlighted the importance of control samples and how conclusions depend on the properties of the population from which those controls are selected.  Specifically, we found that many PSBs appear HI-rich when compared to a stellar mass matched sample (left panel of Figure \ref{delta_fgas}) but have normal-to-low HI masses compared with star-forming galaxies (right panel of Figure \ref{delta_fgas}).  Likewise, if the comparison sample is drawn from the green valley population, PSBs are mostly HI-normal, or slightly gas rich (Figure \ref{dnuvr}).  A similar sensitivity to the choice of control pool was shown by Ellison et al. (2019), where an apparent HI-richness in AGN hosts disappeared when SFR was also accounted for.   The success of any analysis that aims to measure changing properties in a particular population therefore not only requires a large statistical comparison pool from which to construct a control sample, but also careful contemplation of how the matching should be done.

\section{Conclusions}

Post-starburst galaxies represent a powerful tool for studying the shut-down of star formation.  In the work presented here we investigate the role of atomic gas in this quenching process.  By combining new 21cm FAST observations with measurements available in the literature, we have assembled a homogeneous sample that achieves a uniform HI detection threshold for 68 PSBs at $z<0.04$.  Not only is this sample an order of magnitude larger than most previous studies of HI in PSBs, but we have, for the first time, been able to make a statistically rigourous comparison to a control sample of equivalent depth.  Our findings are as follows.

\begin{itemize}

\item  \textbf{PSBs can retain large HI reservoirs.}  We detect HI in 57/68 PSB galaxies with HI masses that range from \mhi\ $\sim10^{8.5}$ up to $10^{10}$ \msun, representing HI gas fractions (\fgas\ = \mhi/M$_{\star}$) from a few percent up to almost 30 percent (Figure \ref{fgas}).  The quenching in these PSB galaxies has therefore clearly been achieved despite the presence of a large atomic gas reservoir.

\item  \textbf{On average, PSBs are a factor of $\sim$2 more HI-poor than their star-forming progenitors.}  When compared to a stellar mass matched-sample drawn from xGASS, the HI gas fractions of PSBs are marginally lower, by $\sim$0.1-0.2 dex, at the lower stellar mass end of the sample (\logm\ $<$ 10.3) and 0.4-0.6 dex larger at higher stellar masses.  However, when compared with only the star-forming galaxies in xGASS (which we argue is a more meaningful comparison, the star-forming galaxies being the presumed progenitor population of PSBs), the median PSB gas fraction is lower by 0.2--0.4 dex at all stellar masses (Figure \ref{med_fgas}).  Therefore, despite retaining an ample atomic gas reservoir, PSBs do exhibit (on average) depleted HI gas fractions compared to star-forming galaxies of the same stellar mass.

\item  \textbf{PSBs show a diversity of HI properties.}  By matching each individual PSB to a bespoke xGASS control sample, we can assess their HI-richness on a case-by-case basis.  Again, we see different behaviour depending on whether we compare to all xGASS galaxies, or only to those that are star-forming.  In the latter case, we find that approximately half of the PSBs are `gas typical', in that their HI masses lie within the expected range of xGASS star-forming galaxies.  The remaining 50 percent are HI-poor, by up to a factor of ten (Figure \ref{delta_fgas}).

\item  \textbf{The HI gas fractions of PSBs are mostly typical of other green valley galaxies, although some are significantly HI-rich.}  In addition to comparing PSBs to their star-forming progenitors, we can also compare them to other galaxies that are in the green valley.   In this way we are able to assess whether their recent and rapid quenching sets them apart from the general population of galaxies with intermediate sSFRs.  By using $NUV-r$ colour as a proxy for sSFR, we find that approximately two thirds of PSBs have HI gas fractions typical of other green valley galaxies, with the final one third being gas-rich by a factor of up to ten (Figure \ref{dnuvr}).

  \medskip

\item \textbf{The HI gas mass of PSBs shows no correlation with time since burst.}  Whereas previous studies have found evidence for a declining molecular gas fraction as a function of time since burst, no such trend is found for the atomic gas content (Figure \ref{bage}).  The persistence of abundant HI reservoirs long after the burst leaves open the possibility for the rekindling of star formation in these galaxies in the future.

\acknowledgements

We are grateful to the anonymous referee whose comments improved several points in this work.  SLE and SW are both grateful for funding from NSERC.  MJJD acknowledges support from the Spanish grant PID2022-138560NB-I00, funded by MCIN/AEI/10.13039/501100011033/FEDER, EU.  JW thanks support of research grants from  Ministry of Science and Technology of the People's Republic of China (NO. 2022YFA1602902), National Science Foundation of China (NO. 12233001), and the China Manned Space Project (No. CMS-CSST-2025-A08). VW acknowledges Science and Technologies Facilities Council (STFC) grant ST/Y00275X/1 and Leverhulme Fellowship RF-2024-589/4.
  
  \appendix

  \section{FAST observed sample}

In Figure \ref{atlas_all} we present the remaining images and spectra for the FAST observed PSB sample.
  
\begin{figure*}
    \centering
    \includegraphics[width=\linewidth]{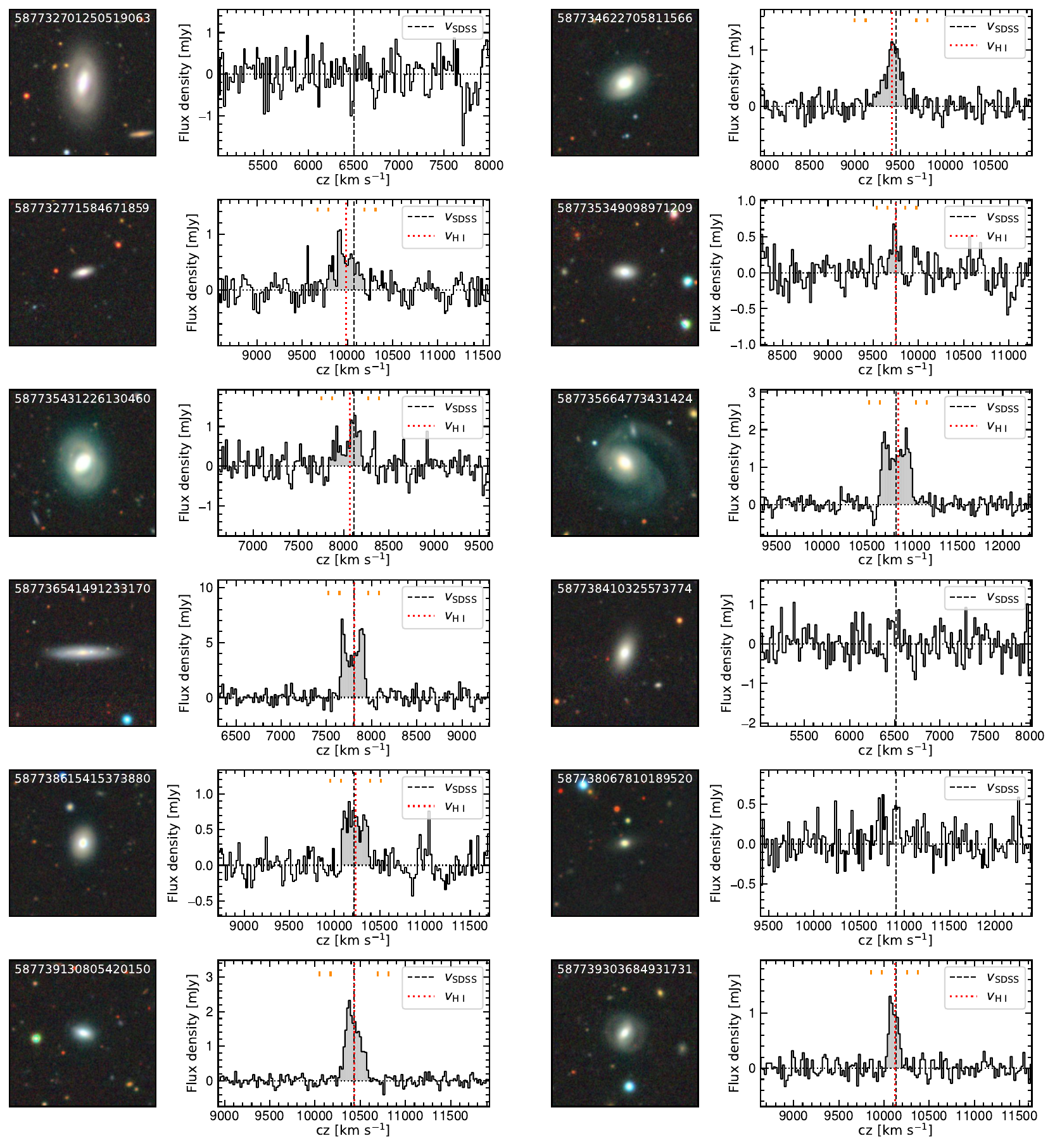}
    \caption{Legacy survey images ($1'.45\times1'.45$, half of the FAST beam size; Dey et al. 2019) and HI spectra for the remaining galaxies in Table \ref{tab:obs}, ordered by the SDSS objID (shown in the images). The HI spectra are baseline-subtracted and binned to a channel width of $20~\rm km~s^{-1}$. The red dotted line and the black dashed line indicate the heliocentric velocity from HI spectra and SDSS, respectively. The orange dashes show the velocity range where the HI growth curve flattens.   }
    \label{atlas_all}
\end{figure*}
 
\begin{figure*}
\addtocounter{figure}{-1}
    \centering
    \includegraphics[width=\linewidth]{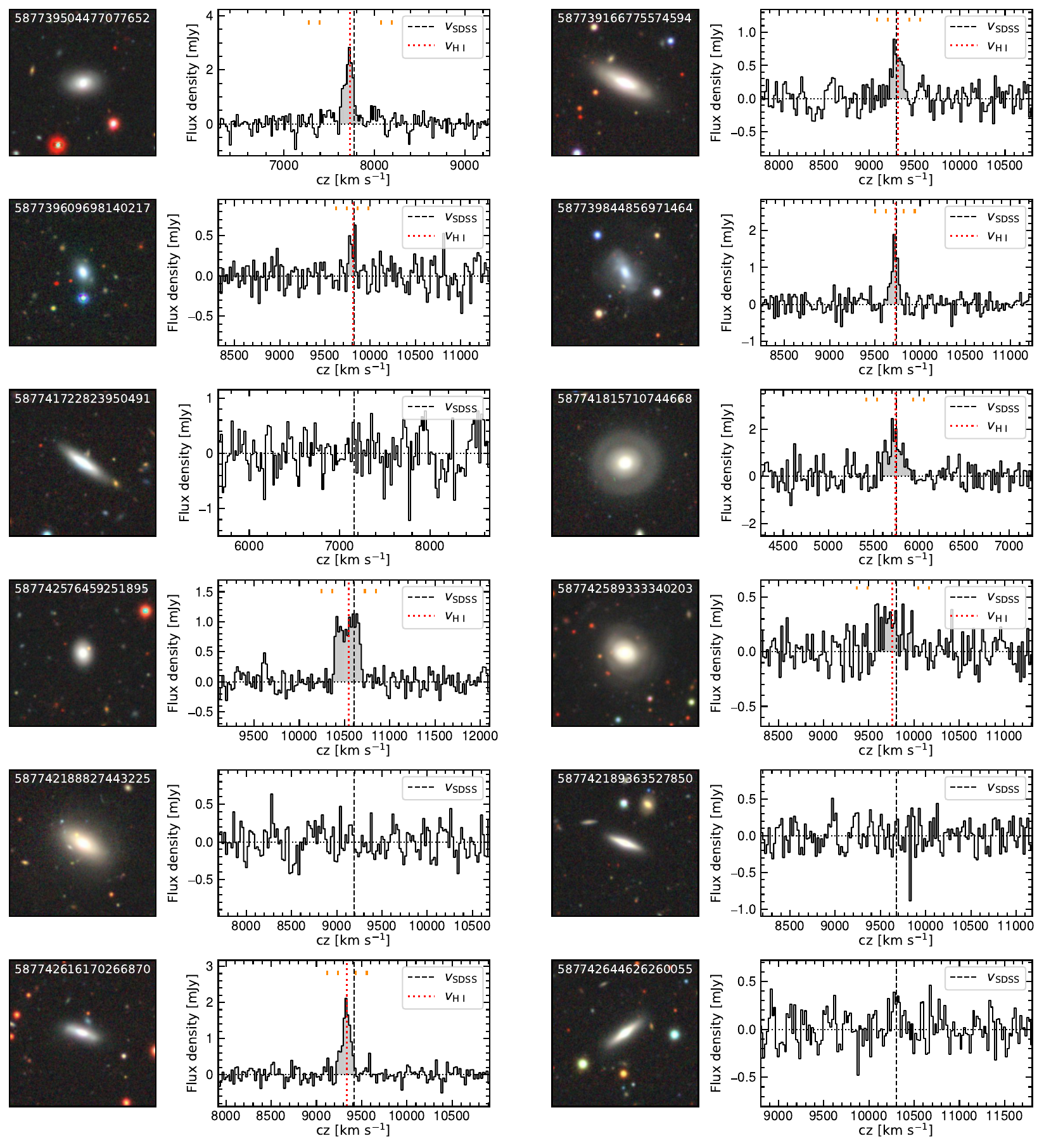}
    \caption{Continued.   }
\end{figure*}
 
\begin{figure*}
\addtocounter{figure}{-1}
    \centering
    \includegraphics[width=\linewidth]{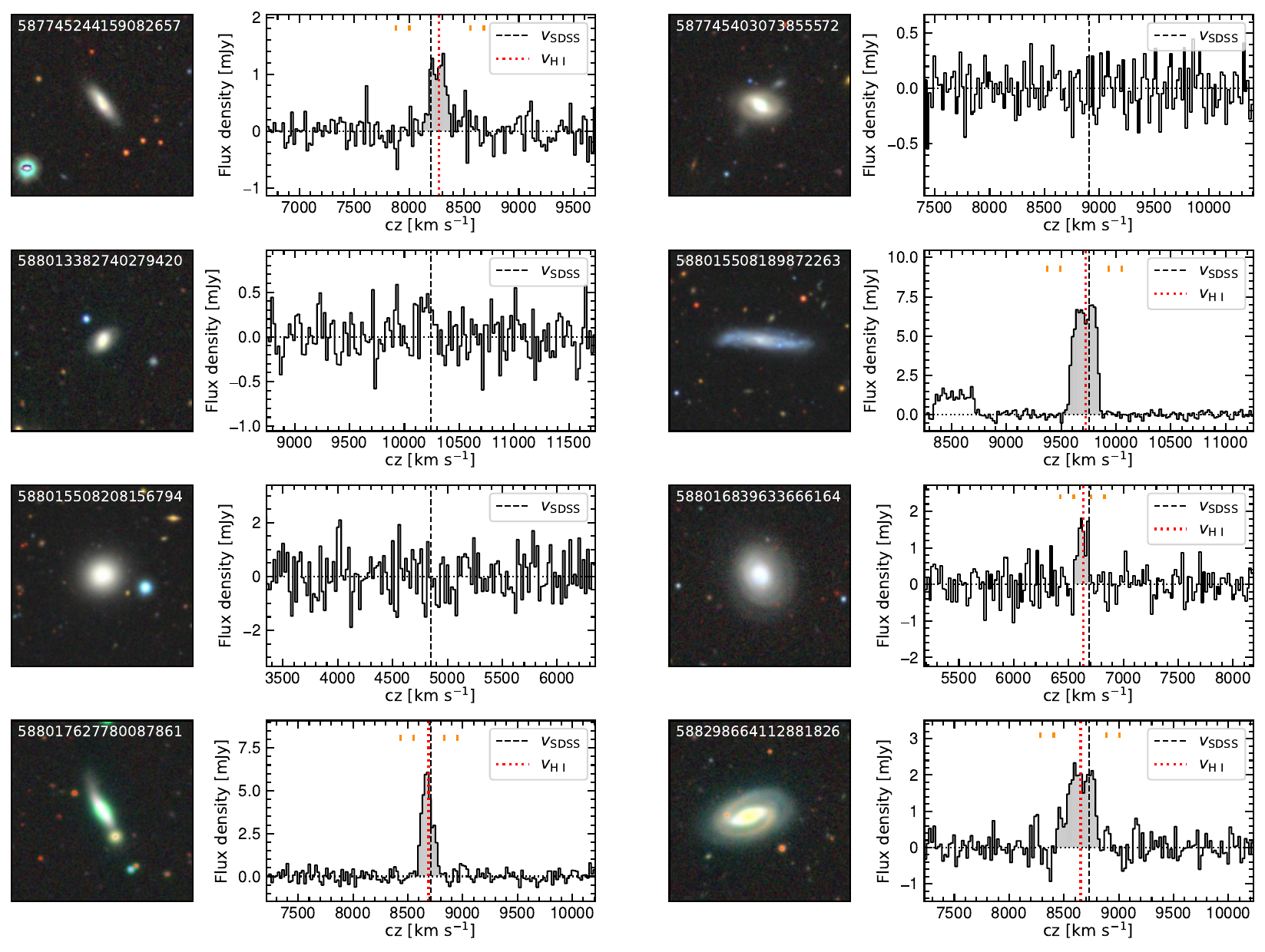}
    \caption{Continued.   }
 \end{figure*}

\end{itemize}

\end{document}